\newif\ifLOWRES 
\LOWRESfalse 
\documentclass[fleqn,usenatbib]{mnras}

\usepackage{newtxtext,newtxmath}

\usepackage[T1]{fontenc}
\usepackage{ae,aecompl}

\usepackage{graphicx}	
\usepackage{amsmath}	
\usepackage{amssymb}	
\usepackage[dvipsnames]{xcolor}
\usepackage{xparse}
\usepackage{xspace}
\usepackage{placeins}

\newcommand\ba{\begin{eqnarray}}
\newcommand\ea{\end{eqnarray}}

\newcommand{\athena}{\texttt{Athena++}\xspace}
\newcommand{\M}{{\ensuremath{\mathcal{M}}}\xspace}

\NewDocumentCommand\pder{mmg}{\ensuremath{
		\IfNoValueTF{#3}
		{\dfrac{\partial #1}{\partial #2}}
		{\left(\dfrac{\partial #1}{\partial #2}\right)_{#3}}
}}

\title[Modes in the Boundary Layer]{Boundary Layers of Accretion Disks: Discovery of Vortex-Driven Modes and Other Waves}

\author[Coleman et al.]{
Matthew S. B. Coleman,$^{1,2}$
Roman R. Rafikov,$^{1,3}$\thanks{Corresponding author: rrr@damtp.cam.ac.uk}\thanks{John N. Bahcall Fellow at the IAS}
Alexander A. Philippov$^{4}$
\\
$^{1}$Institute for Advanced Study, Einstein Drive, Princeton, NJ 08540, USA\\
$^{2}$Department of Astrophysical Sciences, 4 Ivy Lane, Princeton University, Princeton, NJ 08540, USA\\
$^{3}$Centre for Mathematical Sciences, Department of Applied Mathematics and Theoretical Physics, University of Cambridge, \\ \hspace{.05em} Wilberforce Road, Cambridge CB3 0WA, UK\\
$^{4}$Center for Computational Astrophysics, Flatiron Institute, 162 Fifth Avenue, New York, NY 10010, USA
}

\date{Accepted XXX. Received YYY; in original form ZZZ}

\pubyear{2015}

\begin{document}
\label{firstpage}
\pagerange{\pageref{firstpage}--\pageref{lastpage}}
\maketitle

\defcitealias{BR12}{BR12}
\defcitealias{BRS12}{BRS12}
\defcitealias{BRS13a}{BRS13a}
\defcitealias{BRS13b}{BRS13b}
\defcitealias{ARZ}{AR18}

\begin{abstract}
Disk accretion onto weakly magnetized objects possessing a material surface must proceed via the so-called boundary layer (BL) --- a region at the inner edge of the disk, in which the velocity of accreting material abruptly decreases from its Keplerian value. Supersonic shear arising in the BL is known to be conducive to excitation of acoustic waves that propagate into both the accretor and the disk, enabling angular momentum and mass transport across the BL. We carry out a numerical exploration of different wave modes that operate near the BL, focusing on their morphological characteristics in the innermost parts of accretion disk. Using a large suite of simulations covering a broad range of Mach numbers (of the supersonic shear flow in the BL), we provide accurate characterization of the different types of modes, verifying their properties against analytical results, when available. We discover new types of modes, in particular, global spiral density waves launched by vortices forming in the disk near the BL as a result of the Rossby wave instability; this instability is triggered by the vortensity production in that region caused by the nonlinear damping of acoustic waves. Azimuthal wavenumbers of the dominant modes that we observe appear to increase monotonically with the Mach number of the runs, but a particular mix of modes found in a simulation is mildly stochastic. Our results provide a basis for better understanding of the angular momentum and mass transport across the BL as well as the emission variability in accreting objects.         
\end{abstract}





\section{Introduction}
\label{sect:intro}


Accretion disks are ubiquitous in astrophysics, with objects ranging from active galactic nuclei to protostars being fundamentally tied to them. In the cases where the central object (i.e. accretor) is not a black hole, but is a neutron star, a white dwarf, a protostar, or a protoplanet (henceforth we refer to any of these objects as a ``star"), the accretor has a material surface, which the accreted material must connect to in some fashion. If the accretion rate is high and the magnetic field of the star is sufficiently low, then the accretion flow does not get disrupted by magnetic stresses \citep{Ghosh1977,KON91} and the disk can extend all the way to the surface of the star. This particular situation inevitably requires accreting gas to transition from rapid, supersonic rotation (at Keplerian velocity) in the disk to a slow rotation in the star. The region of the disk-star system where this transition takes place is known as the {\it boundary layer} (BL). Systems where the BLs are expected to emerge include e.g. FU Ori type young stellar objects \citep{POP93} and cataclysmic variables (CVs, \citealt{Kip1978,NAR93}). Weakly magnetized neutron stars in low-mass X-ray binaries (LMXBs) are believed to accrete in a broadly similar fashion through the so-called {\it spreading layer}  \citep{INO99,INO10,Gilf2003,Rev2006,PHI16}. Objects accreting gas directly onto their surfaces through the BLs are the subject of the present paper.

In order for the material arriving from the Keplerian disk to become a part of a slowly rotating star it must somehow lose its angular momentum. While the magnetorotational instability (MRI; \citealt{VEL59,Chandra,BAL91}) is traditionally invoked as the favored angular momentum transport mechanism in ionized Keplerian accretion disks, it would not operate in the BL. This is because the MRI requires that the angular frequency $\Omega$ of the fluid flow decays with the distance, $\mathrm{d}\Omega/\mathrm{d}r<0$, whereas the BL naturally has $\mathrm{d}\Omega/\mathrm{d}r>0$, preventing the MRI from operating \citep{Pes2012}. This conclusion has been verified by MHD simulations of the BLs \citep{BRS13b}.

Instead the disk must utilize a different mechanism to remove  angular momentum from the accreting gas, which passes through the BL on its way to the surface of the star. \citet{BR12} identified a robust mechanism for doing that --- a linear instability operating in a supersonic shear flow, which generates acoustic waves in the BL where the azimuthal velocity of the flow exhibits sharp supersonic variation. This instability is global and similar in nature to the Papaloizou-Pringle instability \citep{Drury1979,Drury1980,Drury1985,PP1984,NGG1987,Glatzel1988}. The waves excited by the instability propagate both out in the disk and into the star, allowing energy and angular momentum of accreting gas to be transported over significant distances before being dissipated. Numerical simulations later confirmed that this angular momentum transport mechanism robustly operates within the BL, both in hydrodynamic \citep{BRS12,BRS13a,HER15} and magnetohydrodynamic \citep{BRS13b,BQ17} settings.

This discovery marked a significant paradigm shift compared to the local transport mechanisms invoked in previous studies of the BL problem \citep{Kip1978,POP93,NAR93,HER13}. The intrinsically non-local nature of this mechanism could substantially impact disk thermodynamics and its spectrum \citep{BRS12,BRS13a}. Another important implication follows from the fact that the modes excited in the BL are intrinsically non-axisymmetric. This should lead to the variability of emission produced in the near-BL part of the disk and may explain the various types of quasi-periodic variability observed in objects accreting through the BLs (e.g. CVs, see \citealt{WAR03}).

These ramifications, as well as the ubiquity of the BLs in astrophysics, motivate further efforts to better understand their physics through numerical simulations, building on the previous work of \citet{BRS12,BRS13a,BRS13b}. These past studies, while significantly advancing our understanding of the BL structure, were often limited in terms of the numerical resolution, duration of the simulations, and the number of model parameters that have been varied. 

In this paper, first in a series, we present a new set of long-term, high-resolution, hydro simulations focused on exploring the BL physics. We provide extensive exploration of both the physical and numerical parameter space to test the sensitivity of outcomes to both types of simulation inputs. We carry out an in-depth analysis of the mode structure of the perturbations that arise in the vicinity of the BL as a result of ongoing acoustic instability. A key highlight of this study is the discovery of new types of modes, naturally emerging in this disk region, and our attempts at understanding their origin. In the future we will use this numerical data set to analyze angular momentum and mass transport driven by the different modes operating in the vicinity of the BL (Coleman et al. in prep).

Our paper is organized as follows. In \S\ref{sect:ph-setup} we discuss our physical setup and typical values of the Mach number in different astrophysical objects to motivate the parameter choices for our simulations. We remind the basics of the acoustic mode phenomenology in \S\ref{sect:pheno}, and cover the details of our numerical setup in \S\ref{sect:nmumerics}. We provide detailed morphological description of the various modes that we find in our runs for different values of the Mach number in \S\ref{sect:fiducial} and Appendix \ref{sect:otherMs}. Description of the vortex-driven modes and explanation of their origin are provided in \S\ref{sect:vortex} \& \ref{sect:vortex_disc}, respectively. We discuss properties of other modes found in our simulations, as well as some other aspects of our work, in \S\ref{sect:disc}, and summarize our main findings in \S\ref{sect:sum}.


\section{Physical Setup and Typical Mach Numbers}
\label{sect:ph-setup}


In this work we consider a system consisting of a central object with a surface (a star) and an accretion disk extending all the way to the star, i.e. having a physical contact with its surface. We study the evolution of this system in two-dimensional (2D, vertically integrated), hydrodynamic (i.e. no magnetic fields) setup. The disk is non-self-gravitating and orbits in a Newtonian potential of a central point mass $M_\star$. Very importantly, the disk has no intrinsic viscosity\footnote{Our simulations have no explicit viscosity and numerical viscosity is negligible.} so that any mass re-distribution (accretion) in the system can take place only due to the action of the waves propagating in the disk and the star. 

Similar to a number of previous studies of the BL \citep{BRS12,BRS13a,BRS13b}, we treat disk thermodynamics using the globally isothermal equation of state (EoS), $P=\Sigma c_{s}^2$, where $P$ is the vertically integrated pressure,  $\Sigma$ is the surface density, and $c_{\rm s}$ is the sound speed which is constant. The advantage of using this simple EoS comes from the fact that it keeps the disk thermodynamics unchanged in the course of simulation. It also naturally allows conservation of the angular momentum flux of the waves propagating in the disk \citep{Miranda-ALMA,Miranda-Cooling}, which is very important for us since we are focusing on the wave-driven evolution of the system. Another commonly used EoS, locally isothermal, has been demonstrated by \citet{Miranda-ALMA,Miranda-Cooling} to not conserve the angular momentum of the waves, a phenomenon that would have greatly complicated understanding of the wave-driven transport (Coleman et al. in prep). And an adiabatic EoS $P\propto \Sigma^\gamma$, with $\gamma\neq1$ would lead to evolution of the thermal state of the disk as a result of entropy production at the shocks inevitably arising in the system (see below), again complicating the interpretation of the results. Implicit in our choice of the EoS is the assumption of gas pressure dominating the total pressure, i.e. radiation pressure is neglected. Therefore, results of our study are not directly applicable to accreting neutron stars, for which radiation pressure plays an important role. 

Another reason for using the globally isothermal EoS is that it reduces the number of parameters needed to characterize the system: all details of the disk thermodynamics get captured in a single variable --- constant gas sound speed $c_{\rm s}$. As a result, the key physical parameter governing the behavior of the system in our runs is the dimensionless {\it Mach number} $\M$ defined as the ratio of the Keplerian speed at the surface of the star ($r=R_\star$) to the sound speed $c_s$:
\begin{align}
\label{eqn:mach}
\mathcal{M}&\equiv\frac{\Omega_K(R_\star) R_\star}{c_{\rm s}},
\end{align}
where $\Omega_K=(GM_\star/r^3)^{1/2}$ is the Keplerian rotation rate. 

To provide motivation for the values of $\M$ explored in this work, we estimate $\M$ in some astrophysical systems that may naturally host BLs. We do this by relating the midplane temperature ($T_{\rm m}$) of an optically thick disk to its optical depth ($\tau>1$), accretion rate ($\dot{M}$), and orbital frequency ($\Omega$) in a standard fashion \citep{SS73}:
\begin{align}
\label{eqn:Tc}
T_{\rm m}^4&=\dfrac{3}{8\pi}\dfrac{\dot{M}\Omega^2}{\sigma_{\rm SB}}\tau,
\end{align}
where $\sigma_{\rm SB}$ is the Stefan-Boltzmann constant. Using this relation to compute the characteristic sound speed in the disk via $c_{\rm s}^2=k_{\rm B}T_{\rm m}/\mu$ (where $k_{\rm B}$ is the Boltzmann constant and $\mu$ is the mean molecular weight) and assuming for simplicity that Eqn. (\ref{eqn:Tc}) holds all the way to the surface of the star (i.e. down to $R=R_\star$), we obtain the following general expression for the characteristic Mach number:
\begin{align}
\mathcal{M}=\left(\dfrac{8\pi}{3}\dfrac{G^3\sigma}{k_{\rm B}^4}\right)^{1/8}M_\star^{3/8}\mu^{1/2}\dot{M}^{-1/8}R_\star^{-1/8}\tau^{-1/8}.
\label{eq:Mach-gen}
\end{align}

One obvious class of astrophysical objects that may feature the BLs are the accreting white dwarfs --- CVs and AM CVn systems. Because of thermal instability in the disk, these systems can exist in two states characterized by high or low values of $\dot M$. In the high-$\dot M$ state one finds for the typical parameters of CVs
\begin{align}
\mathcal{M} = 32 \left(\dfrac{M_\star}{0.6M_\odot}\right)^{3/8} \left(\dfrac{\mu}{0.6}\right)^{1/2} \left(\dot{M}_{-9}\,\dfrac{R_\star}{1.4R_{\earth}}\,\dfrac{\tau}{10^4}\right)^{-1/8},
\label{eq:Mach-CV}
\end{align}
where $\dot{M_i}=\dot{M}/10^i M_\odot$ yr$^{-1}$, while for the AM CVn systems\footnote{AM CVns typically have relatively high accretor masses \citep[see e.g.][]{2007ApJ...666.1174R}. This causes them to have small radii due to the mass-radius relation for white dwarfs.}
\begin{align}
\mathcal{M} = 50 \left(\dfrac{M_\star}{0.9M_\odot}\right)^{3/8} \left(\dfrac{\mu}{1.4}\right)^{1/2} \left(\dot{M}_{-9}\,\dfrac{R_\star}{0.8R_{\earth}}\,\dfrac{\tau}{5\!\times\!10^4}\right)^{-1/8}.
\label{eq:Mach-AMCVn}
\end{align}
The optical depth $\tau$ is estimated from stratified shearing-box simulations in the respective regime\footnote{Note there is a factor of 2 difference here as these papers define $\tau_{\rm tot}$ as twice the midplane $\tau$.}: \citet{HIR14} and \citet{COL16} for CVs and \citet{COL18} for AM CVns. In the low-$\dot M$ states of these systems (when $\dot M$ drops by 2-3 orders of magnitude) $\M$ could go up to $\sim 300$, however, the disk is likely to be disrupted by even a weak stellar magnetic field when $\dot M$ is so low.

For FU Ori stars episodically accreting at high $\dot M$ from the protoplanetary disk we find
\begin{align}
\mathcal{M}=3.9 \left(\dfrac{M}{M_\odot}\right)^{3/8} \left(\dfrac{\mu}{0.6}\right)^{1/2} \left(\dot{M}_{-5}\,\dfrac{R_\star}{2R_\odot}\,\dfrac{\tau}{6\times10^5}\right)^{-1/8},
\label{eq:Mach-FUOri}
\end{align}
where $\tau$ is taken from the simulations of \citet{HIR15}.

Motivated by these estimates, and taking into account the numerical constraints that would permit an efficient parameter space exploration, in this work we focus on exploring the BL physics for the values of $\M$ in the range $5\le \M \le 15$. We run at least one simulation for each integer value of $\M$ in this range, although for several characteristic values of $\M$ ($6, 9, 12$) we run multiple simulations to explore the sensitivity to the initial conditions, resolution, etc.


\section{Acoustic mode phenomenology}
\label{sect:pheno}

\begin{figure}
	\includegraphics[width=\linewidth]{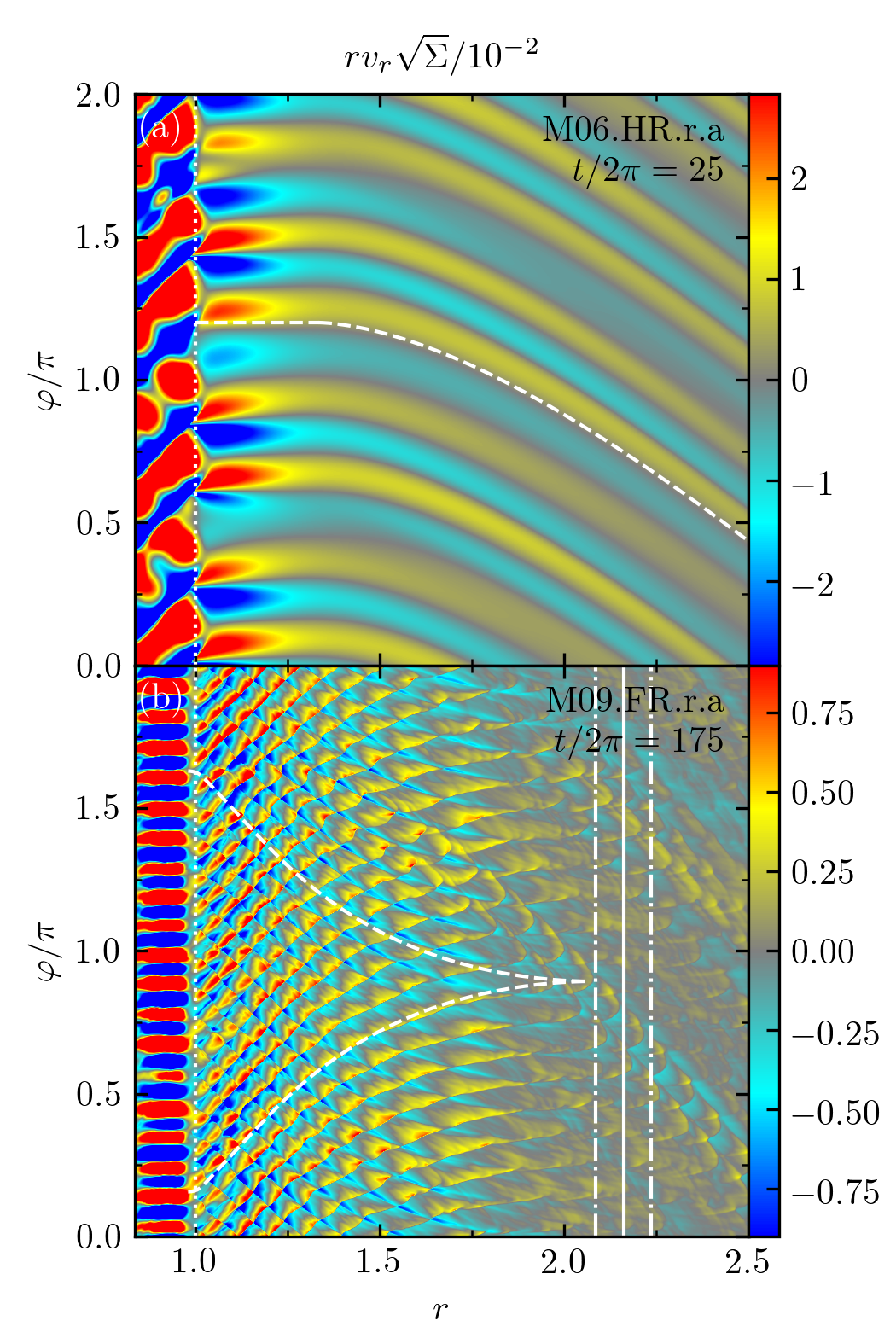}
	\vspace*{-1em}
    \caption{
    Example snapshots from two simulations showing the quantity $rv_r\sqrt{\Sigma}$ (a proxy for acoustic wave action) as a function of $r$ and $\varphi$ in Cartesian coordinates. The white dotted lines at $r=1$ separate the star from the disk. Dashed white curves show the expected shape of the wave pattern for each of the modes, computed using the WKB equations (\ref{eq:disp_rel})-(\ref{eq:wake_shape}) and the measured values of $m$ and $\Omega_{\rm p}$.(a) A snapshot from an $\M=6$ run at 25 inner orbits (during the linear growth phase of the sonic instability) where an $m=7$, $\Omega_{\rm p}=0.758$ global upper mode is clearly visible in the disk (trailing spiral arms) and in the star (inclined wave crests). (b) A snapshot of an $\M=9$ run at 175 inner orbits. The dominant global lower mode with $m=19$ and $\Omega_{\rm p}=0.315$ is clearly visible inside the star and is trapped in the inner disk, $r\lesssim 2$ (Fig.~\ref{fig:M09.FR.r.a-diag} shows $m=19$ as having the most power at $r=0.92,\,1.20$ at this time).  The dot-dashed white lines near $r\approx2.2$ are the inner and outer Lindblad resonances for the $m=19$ mode, while the solid line indicates the corotation radius. The criss-cross structure results from the outward-propagating $m=19$ mode reflecting off the Inner Lindblad Resonance and self-intersecting. }
    \label{fig:sonic-examples}
\end{figure}

We now remind the reader some basic facts about the acoustic modes excited in the BLs. Here we simply summarize the main points made in \citet{BR12} and \citet{BRS12,BRS13a}, which will facilitate the description of the modes found in this work.

Highly supersonic shear present in the BL efficiently drives the non-axisymmetric acoustic (sonic) waves propagating on both sides of the shear layer, in which the azimuthal velocity drop takes place \citep{BR12}. These waves are global and propagate both in the star and in the disk. In general, there are three different types of modes that can get excited in the system, but the simulations typically exhibit only two of them, termed {\it lower} and {\it upper} modes in \citet{BRS13a}. These modes have quite distinct appearance both in the disk and inside the star, and obey very different dispersion relations. They are described in more detail next and are illustrated in Figure \ref{fig:sonic-examples} showing the 2D snapshots of $r v_r \sqrt{\Sigma}$ --- a quantity that should be conserved for a sound wave propagating through the disk. We will routinely display the spatial maps of $r v_r \sqrt{\Sigma}$ in the $r-\varphi$ coordinate plane (with $\varphi$ as the vertical axis), to highlight the details of the morphological features of acoustic modes near $r=R_\star$ where they are excited.


\subsection{Upper modes}
\label{sect:upper}


Upper modes have $k_r\neq 0$ inside the star, as the wave crests of the perturbation pattern associated with this mode are inclined with respect to the radial direction, see Figure \ref{fig:sonic-examples}a. In the disk this mode starts off with $k_r=0$ as $r\to R_\star$ (from above); however, further out in the disk the perturbation pattern gets wrapped up by the differential rotation into multiple trailing spiral arms, see \S\ref{sect:propagation}. Note the sign change of the perturbation variable ($v_r$ in this case) across the BL.

\citet{BRS13a} came up with the following dispersion relation between the azimuthal wavenumber $m$ and the pattern speed $\Omega_{\rm p}$ for the upper modes, which should hold approximately for $m\gg 1$:
\ba  
\left[\Omega(R_\star)-\Omega_{\rm p}\right]^2=\frac{c_s^2}{R_\star^2}+\frac{\kappa^2(R_\star)}{m^2},
\label{eq:DR-uppr-approx}
\ea  
where $\Omega(R_\star)$ and $\kappa(R_\star)$ are the values of the disk angular and epicyclic frequencies as $r\to R_\star$. 
Note that in our runs we often find deviations of $\Omega(r)$ from the Keplerian frequency $\Omega_K$ in the disk near the star (but outside the BL). These deviations are caused by the non-trivial contribution of pressure support to the radial momentum balance
\begin{align}
\Omega^2(r)=\Omega_K^2(r)+\frac{1}{\Sigma r}\frac{\partial P}{\partial r},
\label{eq:Omega}
\end{align}
enabled by the restructuring of the disk surface density near the star (see \S\ref{sect:fiducialM=9}). For that reason, epicyclic frequency $\kappa(r)$ in the disk near the star is generally not equal to $\Omega_K(r)$, as it would be in a purely Keplerian flow.

\begin{figure*}
\includegraphics[width=0.95\linewidth]{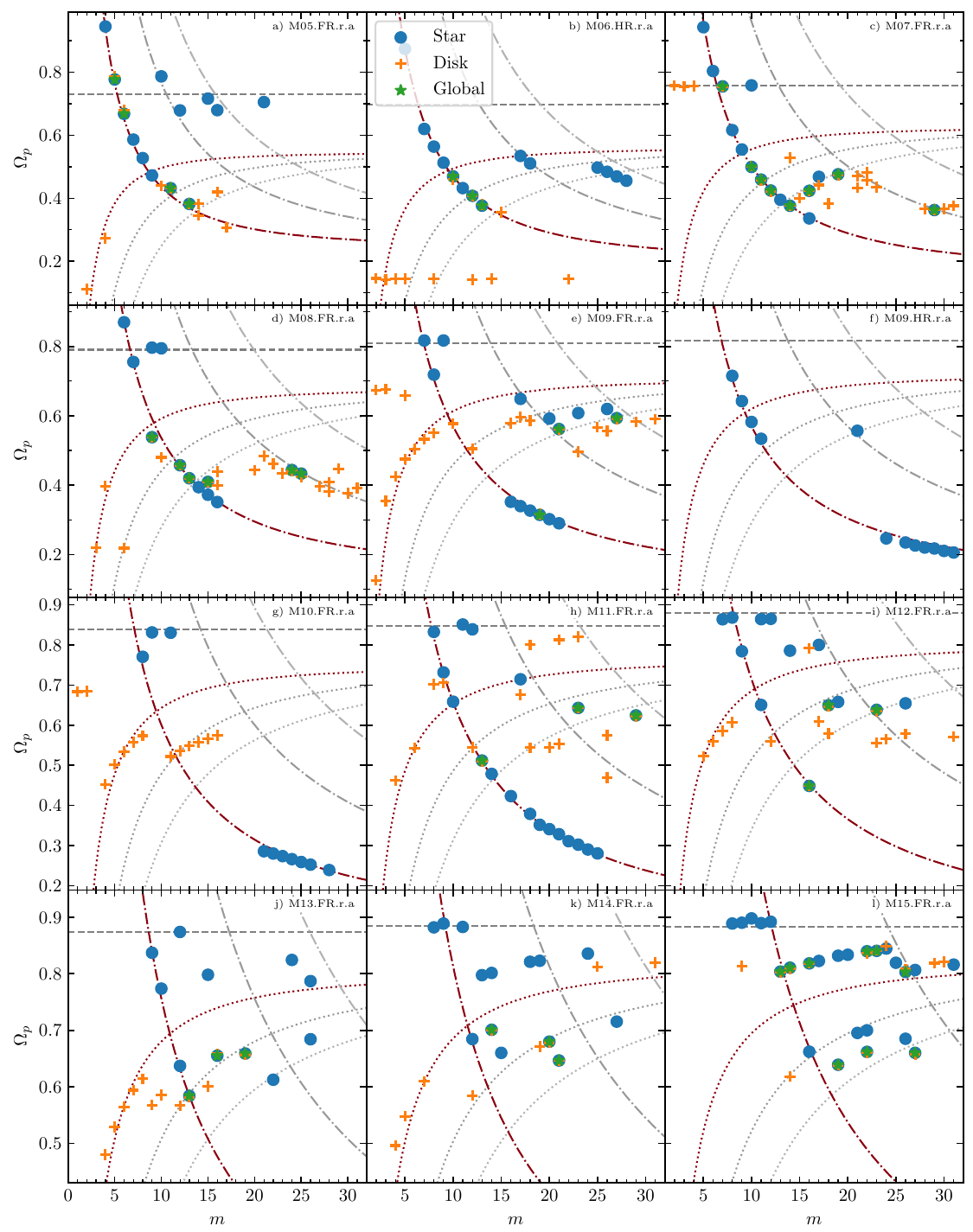}
\caption{Dispersion relation, i.e. a plot of a pattern speed $\Omega_{\rm p}$ vs. the azimuthal wavenumber $m$, for a number of representative simulations with different values of the Mach number $\M$ (labeled in the panels). Runs for all integer values of $\M$ between 5 and 15 are shown, with two runs for $\M=9$ to illustrate the effects of resolution. We show the $(m,\Omega_{\rm p})$ pairs for modes automatically detected in the disk only (at $r=1.2$, yellow pluses), in the star only ($r=(r_{\rm min}+R_\star)/2$, blue circles), and global modes present both in the disk and in the star with the same $m$ and $\Omega_{\rm p}$ at a given time (green stars). Red dotted and dot-dashed curves display the dispersion relations for the upper (\S\ref{sect:upper}) and lower (\S\ref{sect:lower}) modes, correspondingly; grey dotted and dot-dashed curves are their respective higher-order azimuthal harmonics. Horizontal dashed lines show the maximum value of $\Omega(r)$ in the disk at late times. See \S\ref{sect:acoustic-DR} and other text for details. }
\label{fig:multi_dispersion}
\end{figure*}

In this work we analyze our simulation outputs using a relation between $m$ and $\Omega_{\rm p}$, which is more accurate than the Eqn. (\ref{eq:DR-uppr-approx}). The derivation of this refined dispersion relation (\ref{eqn:upper_disp1}) can be found in Appendix \ref{sect:upper-DR} and it is plotted for many values of $\M$ in Figure \ref{fig:multi_dispersion}, clearly showing that $\Omega_{\rm p}$ of the upper modes {\it increases} with increasing $m$. 

\subsection{Lower modes}
\label{sect:lower}


The lower modes exhibit the $k_r$ behavior, which is opposite to that of the upper modes: inside the star they propagate in the azimuthal direction only, i.e. $k_r=0$ for $r<R_\star$. This is illustrated in Figure \ref{fig:sonic-examples}b, where inside the star one can see the perturbation pattern perfectly aligned with the radial direction. At the same time, just outside the star the lower modes have $k_r\neq 0$, and the wave crests are inclined with respect to the radial direction already at $r=R_\star$. Further in the disk $k_r$ gets modified by the differential rotation.

The dispersion relation for the lower modes was derived in \citet{BRS13a} who showed that the pattern speed $\Omega_{\rm p}$ of a lower mode is related to its azimuthal wavenumber $m$ through the following relation
\begin{align}
    \dfrac{\Omega_{\rm p}}{\Omega\left(R_\star\right)}=\dfrac{R_\star}{r_0}\sqrt{\mathcal{M}^{-2}+\left(\dfrac{\mathcal{M}R_\star}{2mr_0}\right)^2},
    \label{eq:lower-DR}
\end{align}
where $r_0$ is a parameter close to unity, see \citet{BRS13a}. Figure \ref{fig:multi_dispersion} illustrates this dispersion relation for a number of values of $\M$, demonstrating that $\Omega_{\rm p}$ of the lower modes {\it decreases} with increasing $m$.


\subsection{Propagation of the BL-excited acoustic modes in the disk}
\label{sect:propagation}


While the appearance of the upper and lower modes just outside the BL is very different (in terms of their $k_r$), their subsequent propagation in the disk follows the classical behavior of the density waves in differentially rotating disks \citep{BT}. In particular, linear density wave theory for a Keplerian disk predicts that an $m$-th azimuthal harmonic of a perturbed fluid variable $f$ behaving as $f\sim \exp(i\Phi_m)$, where 
$\Phi_m=\int^r k_r(r^\prime)dr^\prime+m\varphi$, obeys the standard WKB dispersion relation \citep{GT80}:
\begin{align}
    m^2\left[\Omega_K(r)-\Omega_{\rm p}\right]^2=\Omega_K^2(r)+c_s^2k_r^2.
    \label{eq:disp_rel}
\end{align}
Wave crests trace the trajectory along which the perturbation phase $\Phi_m$ is constant, so that $d\Phi_m=0$. Using the expression for $\Phi_m$, one finds
\begin{align}
    \frac{dr}{d\varphi}=-\frac{m}{k_r(r)}.
    \label{eq:wake_shape}
\end{align}
Integrating this equation with the $k_r(r)$ determined by the equation (\ref{eq:disp_rel}) one obtains the shape (i.e. the $\varphi(r)$ dependence) of the wave crests of the $m$-th harmonic of the fluid perturbation with the pattern speed $\Omega_{\rm p}$. 

In Figure \ref{fig:sonic-examples}a the white dashed curve shows the analytical prediction for the wave crest location computed using the equations (\ref{eq:disp_rel})-(\ref{eq:wake_shape}) and the values of $\Omega_{\rm p}$ and $m$ measured in the upper mode dominating this snapshot (see captions). One can see the analytical calculation agreeing with the actual wave crest shape extremely well and predicting the mode to form a pattern of spiral arms sheared by the differential rotation and {\it propagating out} to large distances (this mode has a narrow evanescent region near the star where $\Omega(r)>\Omega_{\rm p}$, in which $k_r=0$).

The same calculation done for the values of $\Omega_{\rm p}$ and $m$ characterizing the lower mode dominating in Figure \ref{fig:sonic-examples}b also agrees very well with the shape of the outgoing and the incoming sonic waves. One can see that the lower modes are {\it trapped} in a {\it resonant cavity} extending between the stellar surface and the Inner Lindblad Resonance (ILR), at which $k_r=0$ for that mode. For the modes discussed in this work the radial location $r_\mathrm{ILR}$ of the ILR typically ends up far enough in the disk where $\Omega(r)$ can be well approximated by $\Omega_K$. Then $r_\mathrm{ILR}$ is determined by the condition $\Omega_K(r_\mathrm{ILR})(1-m^{-1})=\Omega_{\rm p}$, so that  
\ba   
r_\mathrm{ILR}=R_\star\left[\frac{\Omega_K(R_\star)}{\Omega_{\rm p}}\frac{m-1}{m}\right]^{2/3}.
\label{eq:r_ILR}
\ea
For $m\gg 1$ the location of ILR is close to the {\it corotation} radius $r_\mathrm{c}=R_\star\left[\Omega_K(R_\star)/\Omega_{\rm p}\right]^{2/3}$, see Figure \ref{fig:sonic-examples}b.


\section{Numerical setup}
\label{sect:nmumerics}


We simulate the BL and its vicinity --- outer layers of a star and inner regions of an accretion disk --- in (vertically integrated) cylindrical geometry, using \athena \citep{athena} to solve the hydrodynamic equations
\begin{align}
	\pder{\Sigma}{t}+\nabla\cdot\left(\Sigma\mathbf{v}\right)&=0,\\
	\pder{\Sigma\mathbf{v}}{t}+\nabla\cdot\left[\Sigma\mathbf{vv}+\Sigma c_s^2\mathbf{I}\right]&=-\Sigma\mathbf{\nabla}\Phi,
\end{align}
where $\mathbf{v}$ is the fluid velocity, $\mathbf{I}$ is the identity tensor, and $\Phi$ is the stellar potential. For all runs we used the HLLE Riemann solver, second-order van Leer time integrator, and second-order piecewise-linear primitive reconstruction.

Our simulation domain extends from $r_{\min}$ to $r_{\rm max}$ in the radial direction and covers full $2\pi$ in azimuthal direction ($\varphi$). Our grid is uniformly spaced in $\varphi$ and logarithmically spaced in $r$ (i.e. $\delta r\propto r$). We choose $r_{\rm max}=4R_\star$, far enough in the disk to ensure that the structures emerging in our runs can fit within the simulation domain (e.g. see \S\ref{sect:one-armed}). The inner boundary inside the star is placed at $r_{\min}$ such that the density contrast $\Sigma\left(r_{\rm min}\right)/\Sigma\left(R_\star\right)=10^7$, where we assume an isothermal hydrostatic atmospheric profile inside the star. This choice of $r_{\rm min}$ was made to minimize the simulations dependence of $r_{\rm min}$; see Section \ref{sect:conv} for more details.

We carry out a detailed analysis of our simulation outputs, both on-the-fly and in post-processing. In particular, we analyze the behavior of the fluid variables in Fourier domain and develop a fully automated procedure for detecting and measuring properties of the various wave-like perturbations present in our simulations. These and other analysis modules are described in more detail in Appendix \ref{sect:num_an}. The ability to not only infer the existence of multiple modes and derive mode wavenumber $m$ and pattern speed $\Omega_{\rm p}$, but to also follow their evolution throughout the full duration of a run is what makes our analysis extremely powerful. It enables us to see important trends and patterns across dozsens of the BL simulations performed for different values of $\M$. Some other details of our numerical setup are described below.


\subsection{Units}
\label{sect:units}


To define simulation units we took the surface of the star as our fiducial location. We chose $r=1$ and $\Sigma=1$ to correspond to this location (at $t=0$), and for the Keplerian velocity at the surface of the star $v_K(R_\star)$ to be unity, making $c_s=\mathcal{M}^{-1}$ and $GM_\star=1$. This choice makes the Keplerian period at the surface of the star $\tau_{\star}=2\pi$; this is why we often state times in the form of $t/2\pi$.


\subsection{Initial and Boundary Conditions}
\label{sect:ic}


To create the initial conditions of our simulations we partitioned the simulation domain into three regions:
star ($r\le 1-\delta/2$), transition ($1-\delta/2<r<1+\delta/2$), disk ($r\ge 1 + \delta/2)$, with $\delta=0.05$. We initialized the star in hydrostatic equilibrium (HSE), $\Sigma=\Sigma_0 \exp\left(\mathcal{M}^2/r\right)$, $\mathbf{v}=0$.
The disk is initialized with $\Sigma=r^{-3/2}$, $\mathbf{v}=r^{-1/2}\hat{\mathbf{\phi}}$, i.e. a pure Keplerian disk neglecting pressure support (in the beginning of a simulation the disk quickly adjusts to an equilibrium state accounting for the radial pressure support). The inner and outer (radial) boundary conditions maintain these initial conditions, for the star and disk respectively. There is a smooth transition between the star and the disk such that $\Sigma$ and $\mathbf{v}$ as well as their first and second derivatives (with respect to $r$) are continuous. 

In order to seed the acoustic instability we add random seed perturbations to the velocity field (in the disk region only). Let $R\in[0,1)$ be a random number picked from a uniform probability distribution function (PDF), and $A=10^{-2}$ be the amplitude of the initial perturbation. To examine any possible dependence associated with this choice we tried four different implementations of random seeds:
\begin{enumerate}
\item Block-random: each \textbf{mesh-block ($32\times32$ cells for all our runs)} has the same series of random numbers $R$ (one per cell) and the initial perturbation is $v_r=A R$. 

\item Block-phase-random: similar to Block-random but \begin{align}
    v_r(r,\varphi)=\dfrac{A}{4} \sum_{n=2,3,5,7} \sin (n\varphi+2\pi R).
\end{align}

\item Globally random: each block receives a different series of random numbers with $v_r=A R$. 

\item Prime modes: only four random numbers are chosen ($R_n$) and 
\begin{align}
    v_r(r,\varphi)=\dfrac{A}{4} \sum_{n=2,3,5,7}\sin(n\varphi+2\pi R_n).
\end{align}
\end{enumerate}
The seed type of each of our simulations can be found in Table~\ref{table:allruns}. The final letter in each simulation name indicates the seed used for the random number generator.


\subsection{Numerical Robustness and Convergence}
\label{sect:conv}


We experimented with varying several numerical parameters to ensure that our simulations are insensitive to these choices. To test convergence of our results (see \S\ref{sect:num_convergence}), for $\mathcal{M}=6,9,12$ we tried doubling and halving the resolution. For a given value of Mach number $\M$ our fiducial resolution is typically chosen so as to keep the radial grid scale relative to the disk scale height $\delta r/h$ roughly constant, with $3\%\lesssim\delta r/h\lesssim6\%$. For $\mathcal{M}=9$ we also tried varying the aspect ratio of the grid cells $N_r/N_\varphi=7/8$ and $2$ and found no noticeable differences with the fiducial choice of $N_r/N_\varphi=1$.

The numerical parameter that impacts our results most is the inner radial extent of the simulation, $r_{\rm min}$. As long as $r_{\rm min}$ is such that $\Sigma\left(r_{\rm min}\right)/\Sigma\left(R_\star\right)\approx 10^{7\pm 1}$, we found that there are no substantial differences in the results of runs with different $r_{\rm min}$. 
However, at density contrasts $\gtrsim 10^8$ tiny numeric fluctuations near $r_{\rm min}$ (likely caused by disagreement between analytic and numerical hydrostatic equilibrium) get amplified by the density contrast, resulting in large amplitude radial oscillations of the star. On the other hand, below density contrasts of $\sim 10^5$ (at $r_{\rm min}$) the acoustic waves generated in the BL and  traveling into the star get prematurely truncated by the edge of the simulation domain.


\subsection{Simulation Improvements}
\label{sect:sim}


While the \athena simulations we run are similar to the \texttt{Athena} simulations presented in \citet{BRS12,BRS13a}, ours differ in a few key ways. 
First, all of our runs extend over full $2\pi$ in $\varphi$ while only a handful of the previous simulations cover this angular extent. 
This is important for properly capturing all non-axisymmetric structures emerging in simulations. Second, our simulations cover a larger radial extent, going out to a maximum radius of $4R_\star$ compared to $2.5R_\star$, as before. We found that this allows us to observe structures that have not been reported previously, see e.g.  \S\ref{sect:one-armed}. Third, while the old simulations used a uniform radial grid, we use a logarithmic radial grid giving us higher effective resolution near the stellar surface, where the acoustic instability operates, for the same number of cells\footnote{For our fiducial resolution $\mathcal{M}=9$ simulations this gives us $\sim 2$ times the radial resolution at $R_\star$ compared to a uniform radial grid with the same number of cells.}. 
Forth, we perform Fourier analysis of the outputs by running fast-Fourier transforms (FFTs) on the fly (see Appendix~\ref{sect:FFT}) giving us new, previously unattainable, diagnostic capabilities. 
These improvements allow us to identify a statistically significant sample of modes emerging in our simulations, and to discover and quantify new types of modes. 

\begin{figure*}
\ifLOWRES
	\includegraphics[width=\linewidth]{figs/lowres/lowres_M09.FR.r.a_maps_stripes_2.png}
\else
    \includegraphics[width=\linewidth]{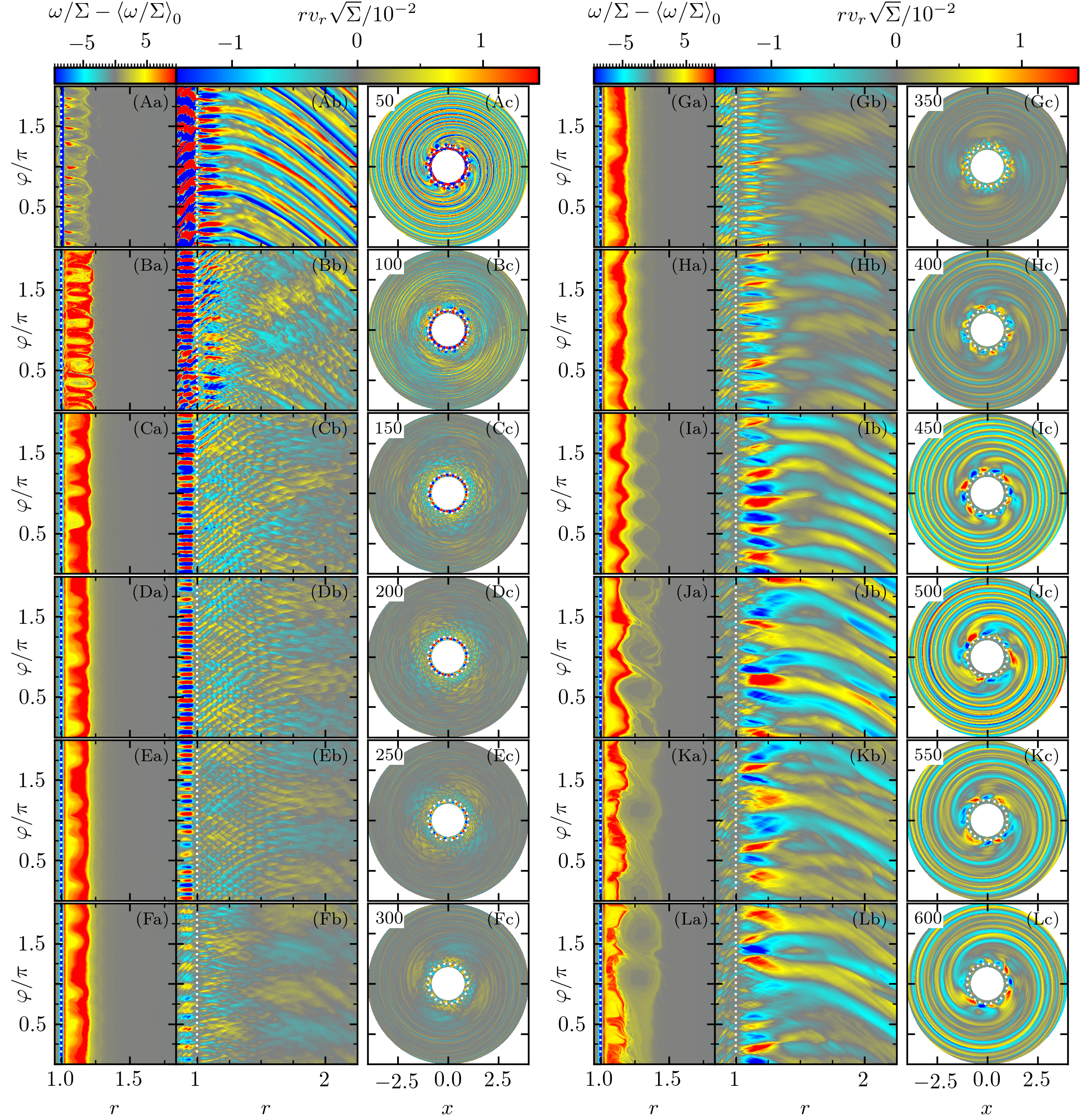}
\fi
    \caption{Maps of ({\it left}) the vortensity perturbation relative to its initial value, ({\it center}) the perturbation measure $r v_r\sqrt{\Sigma}$ in the $r-\varphi$ coordinate plane for a range of $r$, and ({\it right}) $r v_r\sqrt{\Sigma}$ in physical space ($x(r,\varphi),y(r,\varphi)$) in the full simulation domain for $\M=9$ simulation M09.FR.r.a. 
    Different panels correspond to different moments of time labeled in each row. Color bars on top show the scale of the vortensity perturbation and $r v_r\sqrt{\Sigma}$ for this run. See text in \S\ref{sect:fiducialM=9} for detailed description of the evolution shown in this figure.}
    \label{fig:M09.FR.r.a}
\end{figure*}

\begin{figure}
	\includegraphics[width=\linewidth]{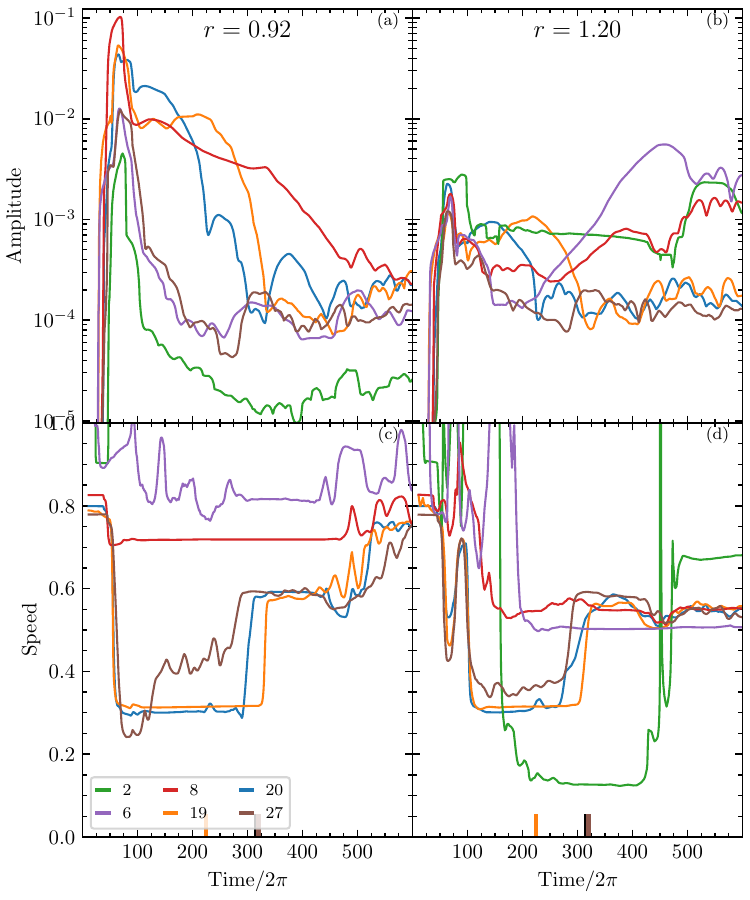}
    \caption{
    Time evolution of the Fourier amplitudes of $v_r\sqrt{\Sigma}$ (top panels) and pattern speeds $\Omega_{\rm p}$ (bottom panels) of a subset of the most prominent modes in the disk-star system for the $\M=9$ run M09.FR.r.a. 
    Different columns illustrate the mode amplitude and $\Omega_{\rm p}$ at different radii: (left) $r=0.92$, inside the star, and (right) at $r=1.2$, in the inner disk. 
    Different curves are color coded according to the azimuthal wave number of the mode that they represent, labeled in the inset. 
    The large ticks on the lower most horizontal axis indicate the temporal midpoint of an automatically detected global mode (see Section \ref{sect:FFT}). If any of these ticks corresponds to one of the plotted modes, then they are drawn in the same color (e.g. the orange tick at $t/2\pi\approx225$).}
    \label{fig:M09.FR.r.a-diag}
\end{figure}

\begin{figure}
	\includegraphics[width=\linewidth]{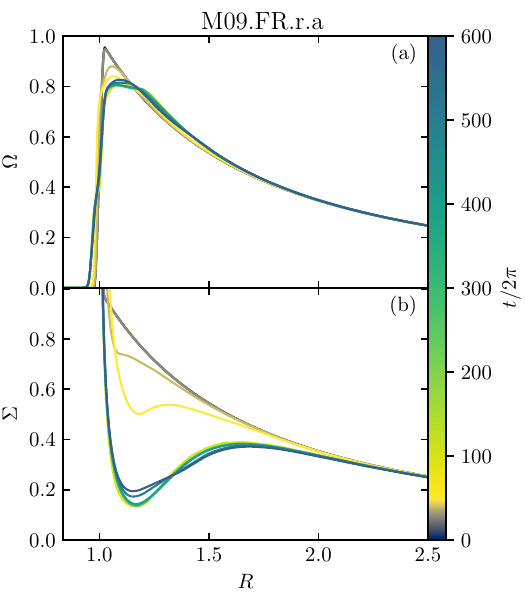}
	\vspace*{-2em}
    \caption{Time evolution of the azimuthally-averaged (a) angular frequency $\Omega(r)$ and (b) surface density $\Sigma(r)$ for the $\M=9$ run M09.FR.r.a described in Section \ref{sect:fiducialM=9}. The colors indicated by the color bar denote the time in inner orbital periods (i.e. $t/2\pi$). See text for details.}
    \label{fig:M09.FR.r.a_evo_prof}
\end{figure}


\section{Morphological characterization of wave modes}
\label{sect:fiducial}


In this section we provide a systematic description of the modes that emerge in a particular $\M=9$ simulation and their evolution, see \S\ref{sect:fiducialM=9}. Simulations with other values of $\M$ are covered in Appendix \ref{sect:otherMs}. Readers not interested in such details can skip these sections. 

Table~\ref{table:allruns} lists the details for the runs in our simulation suite, including their Mach number $\M$, the value of $r_{\rm min}$, resolution and the type of initial noise pattern used to trigger the acoustic instability in the BL (see \S\ref{sect:ic}). To streamline the comparison of runs with different $\M$, in the following we focus on simulations that start with the same initial setup --- a particular realization of the block-random noise pattern, see \S\ref{sect:ic}; for that reason all these simulations have 'r.a' in their label. At the same time, we also provide comparison with simulations using other types of initial conditions (while keeping $\M$ fixed). 

We use several types of diagnostics to illustrate our observations. To highlight the development and subsequent evolution of the wave modes we use the 2D snapshots of $r v_r \sqrt{\Sigma}$ in physical coordinates ($x(r,\varphi),y(r,\varphi)$) at different moments of time starting from the linear development of the instability until the end of the simulation, see the right columns of Figures \ref{fig:M09.FR.r.a}, \ref{fig:M06.HR.r.a}, \ref{fig:M12.FR.r.a}, \ref{fig:M15.FR.r.a}. Also, to highlight the details near $r=1$, we supplement these maps of $r v_r \sqrt{\Sigma}$ with their projections onto $r-\varphi$ coordinate plane, see the central columns of the same figures. Left columns of these figures illustrate the evolution of the flow vortensity, see \S \ref{sect:vortex}.

Harmonic content of the wave modes is illustrated in Figures \ref{fig:M09.FR.r.a-diag}, \ref{fig:M06.HR.r.a-diag}, \ref{fig:M12.FR.r.a-diag}, \ref{fig:M15.FR.r.a-diag} for $\M=9, 6, 12$ and $15$, correspondingly. There we show the amplitudes $A_m$ and pattern speeds $\Omega_{\rm p}$ of the dominant modes (labeled by their $m$) identified by our automated mode detection procedure (see \S\ref{sec:mode_detect} \& \ref{sect:averaging}) as a function of time at two different radii, inside (at $r$ shown in each figure) and outside (at $r=1.2$) the star. This plot allows us to see the transitions between the different types of modes during the simulation. To interpret the nature of the observed modes we will later (see \S\ref{sect:dom_modes}) use the Figure \ref{fig:multi_dispersion}, which displays the different branches of the dispersion relation for upper and lower modes.


\subsection{A typical $\M=9$ run.}
\label{sect:fiducialM=9}


Figure \ref{fig:M09.FR.r.a} illustrates the development and operation of the different modes in an $\M=9$ simulation M09.FR.r.a, which was run at resolution $4096\times 4096$, see Table \ref{table:allruns}. In the beginning of the simulation, sonic instability starts off in the form of an upper $m=27$ mode (not shown in this figure). By $t/2\pi=50$, shown in Figure \ref{fig:M09.FR.r.a}Ab-Ac, the instability reaches saturation with $v_r/c_s \sim$ several per cent; inside the star the upper mode (with $k_r\neq 0$) is already significantly affected by the growing lower $m=9$ mode (with $k_r=0$ for $r<1$).

Outside the star we observe large scale spiral arms extending into the outer disk, reminiscent of the upper mode behavior. However, the number of arms at $r>1$ is not equal to azimuthal wavenumber $m=27$ of the upper mode visible at $r<1$, it is closer to $6$ or $7$. We will discuss the origin of this pattern in \S \ref{sect:vortex}.

By $t/2\pi=100$ shown in Figures \ref{fig:M09.FR.r.a}Bb-Bc, the upper mode weakens considerably (see Fig. \ref{fig:M09.FR.r.a-diag}) and the perturbation pattern is dominated by a superposition of several lower modes (their $k_r\approx 0$ inside the star) with $m=19, 20$. 
One can also see the hints of the emergence of an $m=2$ pattern in the disk, manifesting itself at $t/2\pi=100$ as two broad {\it leading} arms for $r\lesssim 2.5$. The $r-\varphi$ shape of these arms is broadly consistent with what one would expect from an $\Omega_{\rm p}\approx 0.32$ lower mode, suggesting that this low-$m$ pattern may be somehow related to the $m=19,20$ lower modes present in the system. 

These transitions are accompanied by the evolution of disk surface density $\Sigma$ and angular frequency $\Omega$ near the stellar surface, as illustrated in Figure \ref{fig:M09.FR.r.a_evo_prof} at different moments of time. At around $t/2\pi=100$ a number of features start to develop in the $\Omega(r)$ profile in the inner disk, see Figure \ref{fig:M09.FR.r.a_evo_prof}a: an inflection point-like transition at $\Omega\approx 0.4$ inside the BL, a plateau for $1\lesssim r\lesssim 1.2$, and a slightly super-Keplerian rotation for $r\gtrsim 1.2$.  All these features are caused by accretion of gas from the disk (driven by the dissipation of acoustic modes) onto the star, which is revealed by the reduction of $\Sigma(r)$ compared to its initial profile for $r\lesssim 1.7$, see Figure \ref{fig:M09.FR.r.a_evo_prof}b. This depletion, or gap, is quite substantial near $r=1$ ($\Sigma$ drops to $20-30\%$ of its initial value at $r=1.2$) and severely modifies the radial pressure support in this part of the disk. In agreement with the equation (\ref{eq:Omega}), this has a direct impact on the $\Omega(r)$ behavior: $\Omega(r)$ develops a sub-Keplerian plateau in the part of the gap where $\Sigma(r)$ decreases with $r$ (i.e. $r\lesssim 1.2$), and becomes slightly super-Keplerian outside of this region, since $\Sigma(r)$ increases over a range of $r$ there. 
These features will be discussed in more details and across different values of $\M$ in Coleman et al. (in prep.).

Beyond $\sim 100$ orbits the system settles into a less chaotic state (amplitude of $v_r$ variations decreases by $\sim 2-4$ to $v_r/c_s<10^{-2}$), which persists until about $t/2\pi=300$, see Figure \ref{fig:M09.FR.r.a}C-F. During this time the prominent $m=19$ lower mode becomes quite coherent both in the disk and inside the star (see Figure \ref{fig:M09.FR.r.a-diag}a,b). Its relatively low pattern speed $\Omega_{\rm p}=0.315$ corresponds to the corotation radius $r_c=\Omega_{\rm p}^{-2/3}\approx 2.16$ (assuming a Keplerian rotation curve, which is a good assumption at these radii), in good agreement with the radius at which the outwardly propagating wave crests reach $k_r=0$, turn around, and start propagating inwards, towards the star. Thus, $m=19$ mode becomes {\it trapped} inside the radially extended region --- the \emph{resonant cavity} --- between the stellar surface $r=1$ and the inner Lindblad resonance which is close to $r_c$. The interference of the outward/inward propagating waves at $1<r\lesssim 2$ gives rise to a regular pattern of criss-crossing leading and trailing spiral arms confined to the resonant cavity and rotating with a fixed angular frequency $\Omega_{\rm p}$ on top of the (largely) Keplerian flow in the disk, see \S\ref{sect:propagation} and \citet{BRS12,BRS13a}.

During the same period, the aforementioned $m=2$ mode grows in intensity and very noticeably changes its morphology: it turns into a {\it radially elongated}, azimuthally extended perturbation pattern that undergoes a phase shift by $\pi$ at around $r=1.4$. This mode has very low pattern speed  $\Omega_{\rm p}\approx 0.15$ putting its corotation radius at $r_c\approx 3.5$, still inside our simulation domain. We discuss this mode in more detail in \S\ref{sect:krzero_disc}, but note here that it persists until about $400$ orbits, co-existing with the other modes produced at the BL.

Around $300$ orbits the significance of the previously dominant lower $m=19$ mode goes down both inside and outside the star; by $t/2\pi=350$ the associated regular criss-crossing pattern inside the resonant cavity essentially disappears. Simultaneously, an upper $m=23$ mode starts emerging inside the star, with $k_r\neq 0$ for $r<1$.
Interestingly, outside the star our data do not show this $m=23$ mode: we do see strong spiral arms with $k_r=0$ near the star and extending all the way into the disk, but there are few of them, only 5 or 6, instead of $m=23$ as would be appropriate for the global upper mode (which certainly exists inside the star). 
As time goes by, the number of these global spiral arms in the disk (at $r\gtrsim 1.3$) decreases, as if they were merging together, and after $500$ orbits only 2 or 3 of them remain in the disk, somewhat chaotic in appearance. Such global, low-$m$ spirals are seen in a number of our runs and represent a novel feature of the BL simulations that will be discussed in more details in \S\ref{sect:vortex},\S\ref{sect:vortex_disc}. 

Also, starting at around $t/2\pi\approx 400$, a strong $m=6$ perturbation pattern, radially confined within $1<r\lesssim 1.25$, develops in the disk. It is most coherent around $t/2\pi\approx 450$, but can be easily traced until the end of the run (using our automated mode detection algorithm), interfering with the other modes operating in the system. The nature of this perturbation will be discussed in \S\ref{sect:low-m_disc}.


\section{Discovery of vortex-driven modes in the near-BL region}
\label{sect:vortex}


In addition to spatial distributions of $rv_r\sqrt{\Sigma}$, which illustrate the amplitude of the wave-like perturbation, we also examined the maps of vortensity (or potential vorticity, related to the vorticity $\mathbf{\omega}\equiv\mathbf{\nabla}\times \mathbf{v}$)
\ba  
\zeta\equiv\dfrac{\mathbf{\omega}}{\Sigma}=\dfrac{\mathbf{\nabla}\times \mathbf{v}}{\Sigma},
\label{eq:vortensity}
\ea  
which are shown in the $r-\varphi$ coordinate plane in Figures \ref{fig:M09.FR.r.a}, \ref{fig:M06.HR.r.a}, \ref{fig:M12.FR.r.a}, \ref{fig:M15.FR.r.a} (left columns). 

\begin{figure}
\ifLOWRES
	\includegraphics[width=\linewidth]{figs/lowres/lowres_vortex_types.png}
\else
    \includegraphics[width=\linewidth]{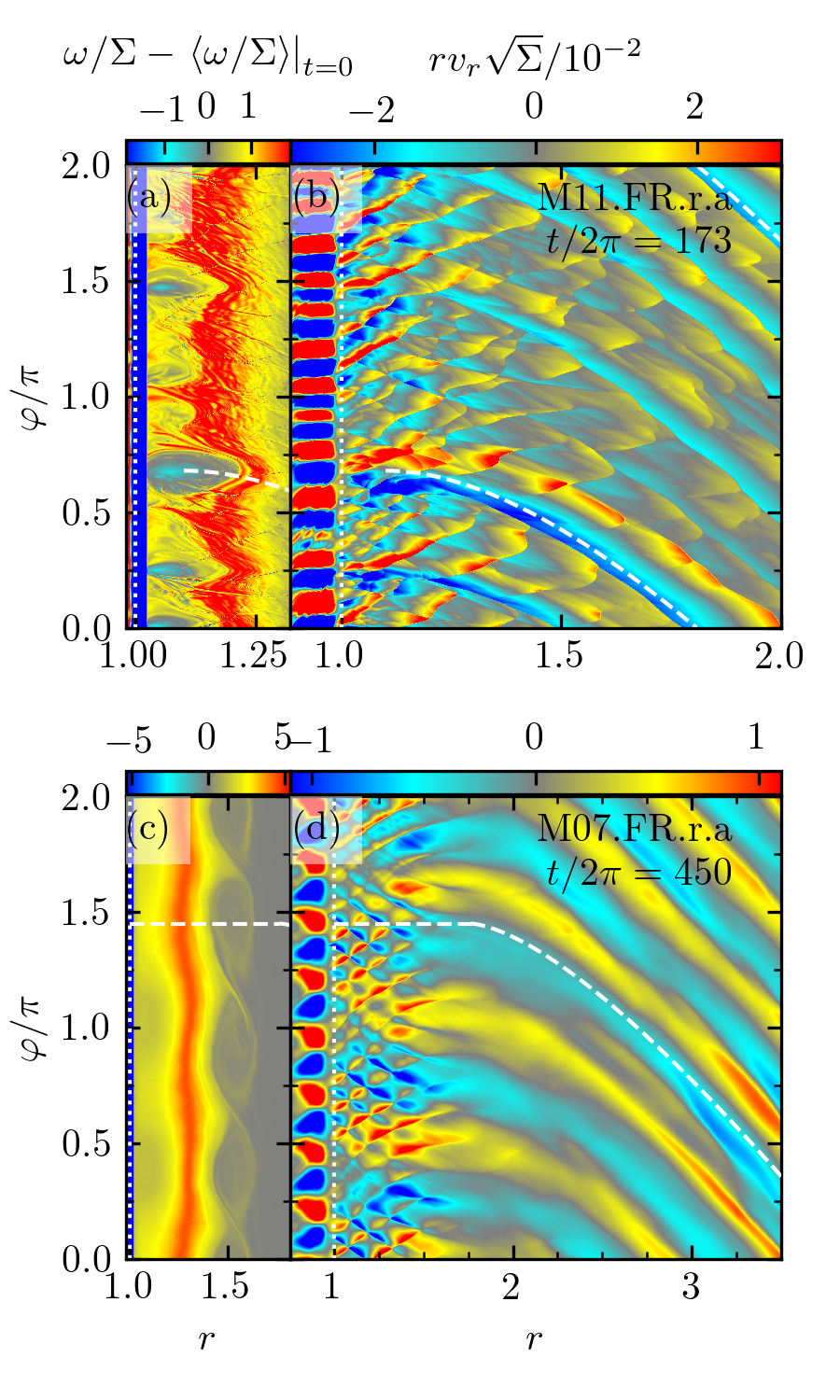}
\fi
	\vspace*{-1em}
    \caption{
    Two main types of localized vortensity structures (vortensity change relative to its initial value) emerging in our simulations (left) and the fluid perturbation $r v_r\sqrt{\Sigma}$ that they drive in the disk (right). (a,b) Compact vortices present close to the BL, around $r\approx 1.1$ in the $\M=11$ run M11.FR.r.a at $t/2\pi=173$. (c.d) Regular ``rolls" forming in the inner disk around $r\approx 1.5$ in the M07.FR.r.a run at $t/2\pi=450$. White dashed curves (on the right) represent the locations of the wave crests according to the WKB dispersion relation (\ref{eq:disp_rel}), and their association with the corresponding vortex structures (on the left). See \S\ref{sect:vortex} for further discussion.
    }
    \label{fig:vortex-types}
\end{figure}

These maps reveal that many of the morphological structures observed in our simulations and mentioned in \S \ref{sect:fiducial} are, in fact, caused by the localized structures in the spatial distribution of $\zeta$ that emerge in the near-BL region. Quite generally, we find two types of vortensity structures that give rise to global waves in disks. Their typical appearance is illustrated in Figure \ref{fig:vortex-types}, where we plot both vortensity and $rv_r\sqrt{\Sigma}$ for a couple of representative runs. 
 
The first type of vortensity structures reveals itself in $\zeta$ map in Figure \ref{fig:vortex-types}a (showing a snapshot of $\M=11$ run at 175 orbits) as sharp, elliptical, anticyclonic features located very close to the stellar surface. We call these structures simply {\it vortices}. They appear rather narrow in azimuthal direction but this is simply a result of the aspect ratio chosen in this figure --- in reality they are rather elongated in $\varphi$. Nevertheless, these vortices are typically well-isolated in azimuthal direction while sharing the same radial range $1<r\lesssim 1.2$ like beads on a wire. 
Looking now at Figure \ref{fig:vortex-types}b, one immediately notices that azimuthal positions of these vortices coincide extremely well with the starting azimuthal locations (at $r\approx 1$) of a number of sharp, narrow spiral arms that propagate out into the inner disk. The strength of the arm (amplitude of its $rv_r\sqrt{\Sigma}$) appears to scale with the size of the vortex to which the arm in connected. The number of arms --- about 7 --- is the same as the number of noticeable vortices in panel (a) of the figure. One can also see that the global spirals in the disk co-exist with the global lower $m=13$ acoustic mode (easily visible inside the star) --- a very different type of the wave-like perturbation. 

Second type of vortensity structure is illustrated in Figure \ref{fig:vortex-types}c, which shows a snapshot of the $\M=7$ run at 450 orbits. This vortensity map reveals a set of four azimuthally elongated ``rolls", as we call\footnote{We often collectively refer to isolated vortices and rolls as just ``vortices".} these structures, which are centered at $r\approx 1.5$ and have approximately equal azimuthal extent. Unlike vortices, the rolls are not isolated and touch each other, collectively covering the full circumference of the disk. Another difference with respect to vortices is that the rolls are always found at some separation from the stellar surface; in Figure \ref{fig:vortex-types}c they occupy a radial range $1.4\lesssim r\lesssim 1.6$.   

Comparing panels (c) and (d) of Figure \ref{fig:vortex-types} one can see that each roll is associated with a broad spiral arm easily visible in $rv_r\sqrt{\Sigma}$. Just as the rolls, the spiral arms are azimuthally broad, which distinguishes them from the narrow arms launched by the vortices. This results in a distinct $m=4$ pattern of global spirals in the outer disk. The leading half of each roll is connected to the $v_r>0$ part of the corresponding spiral arm, while the opposite is true for the trailing half of the roll, indicating their anticyclonic nature (same as vortices). Also, Figure \ref{fig:vortex-types}b shows that the roll-driven spirals arms can naturally co-exist with the acoustic modes, in that case a lower mode which is rather strong in the disk out to $r\approx 1.75$.

Since the starting points of the global spirals (their azimuthal locations at $r\to 1$) always coincide with the positions of their associated vortices/rolls, the pattern speed of the spiral arms in our runs is the same as the orbital frequency of these vortensity structures. As both vortices and rolls are passively advected with the fluid, angular frequency $\Omega(r)$ of the disk fluid at their orbital radii sets $\Omega_{\rm p}$ of their global spirals. Given that rolls are more distant from the stellar surface than the vortices, $\Omega_{\rm p}$ of the spirals associated with vortices is higher than $\Omega_{\rm p}$ of the spirals related to rolls. 

The two kinds of vortensity structures described above emerge at different times in many (but not all) of our simulations, and can even co-exist for brief periods of time. Moreover they tend to evolve and exhibit transitions from one type of structure to another. We now briefly describe the typical evolutionary patterns of vortex-driven modes in runs with different $\M$.


\subsection{Vortensity structures in $\M=9$ runs}
\label{sect:vortex=9}


Our fiducial $\M=9$ run illustrated in Figure \ref{fig:M09.FR.r.a} features a set of isolated vortices emerging near the stellar surface ($r\lesssim 1.25$) by $t/2\pi=50$. These vortices are the true reason behind a set of strong global spirals that are visible in panel (Ab) of this figure (and not the upper mode, as mentioned in \S \ref{sect:fiducialM=9}). They persist at 100 orbits, and their associated spiral arms are discernible in the disk even in the face of a strong lower mode that develops in the system. However, beyond that point vortices merge with each other and get washed out. Correspondingly, the characteristic narrow spiral arms in the disk disappear leaving only the lower mode. 

Beyond 400 orbits a new transition takes place in the system --- a set of rolls starts to emerge at $r\approx 1.3$. At $t/2\pi=450$ one can see 5 regular, roughly equally spaced rolls connected to a set of 5 strong global spirals in the disk. These rolls evolve by merging with each other: only 3 of them remain at 500 orbits (still at roughly equal azimuthal separation from each other), connected to an $m=3$ set of global spirals in the disk, see panels (Ja)-(Jc). Only 2 rolls (and spirals) remain at 550 orbits, separated by roughly $180^\circ$. However, by 600 orbits they drift azimuthally towards one another (while remaining at the same radial distance) and would merge into a single roll if we ran this $\M=9$ simulation for longer.


\subsection{Vortensity structures in the high-$\M$ runs}
\label{sect:vortex>9}


At higher values of $\M>9$ we typically find rolls to emerge quite early. For example, in Figure \ref{fig:M12.FR.r.a}Aa illustrating an $\M=12$ run described in \S \ref{sect:M=12}, a number (7 or 8) of rolls become apparent at $r\approx 1.17$ already at 50 orbits, when a number (9 or 10) of strong, isolated vortices is still present closer to the star. Careful examination of the panel (Ab) of that figure reveals two complexes of global spirals --- one due to the vortices next to the BL and another one associated with the rolls, forming further out in the disk. They can be distinguished by their different pitch angles: roll-driven spiral have lower $\Omega_{\rm p}$ and are less tightly wound than the vortex-driven spiral arms, which have higher $\Omega_{\rm p}$. Because of the difference of their $\Omega_{\rm p}$, the two sets of spirals drift azimuthally relative to each other.  

Co-existence of rolls and vortices persists in this $\M=12$ run for quite a while, with both types of structures (and their associated spirals) visible up to $250$ orbits. However, the number of both vortensity structures goes down as they merge, while maintaining roughly the same radial distance. Vortices near the stellar surface stop being visible only after $\approx 300$ orbits, see panel (Da). 

In this particular simulation vortensity distribution also tends to develop a banded structure after about 150 orbits. Radially narrow and almost azmuthally symmetric bands in $\zeta$ maps appear to give rise to weaker rolls at larger separation from the star. This complicated radial distribution of $\zeta$ goes away only at the end of the simulation, although the rolls at $r\approx 1.1$ still persist in some form. 

A similar evolution of vortensity structures is found in the $\M=15$ run described in \S \ref{sect:M=15}. Left row of Figure \ref{fig:M15.FR.r.a} shows strong vortices early on (panel Aa), which co-exist with a number of rolls later on (panels Ba and Ca), with rolls dominating after $\approx 300$ orbits. These vortensity structures explain the global spirals visible in the maps of $rv_r\sqrt{\Sigma}$ at various degree of coherence throughout the $\M=15$ run.


\subsection{Vortensity structures in the low-$\M$ runs}
\label{sect:vortex<9}


Situation is quite different in our runs with low values of $\M<9$. We find that only the $\M=7$ run shows the development of strong vortices and, subsequently, rolls, reminiscent of the $\M=9$ run; similarity of the perturbation morphology between the $\M=7$ and 9 runs has been previously noted in \S \ref{sect:M=6}. On the other hand, $\M=8$ run does not show any strong or long-lasting azimuthal vortensity structures --- the distribution of $\zeta$ in this run looks quite axisymmetric throughout its duration. And the simulations with $\M=5$ and 6 develop rather peculiar vortensity structure, illustrated in the left column of Figure \ref{fig:M06.HR.r.a}, which is very distinct from the higher $\M$ runs. 

The $\M=6$ run M06.HR.r.lc.a shows near-stellar surface vortices only for a very brief interval of time around 75 orbits (not shown). And soon after a strong $m=2$, low-$\Omega_{\rm p}$ mode (described in \S \ref{sect:M=6}) appears in the disk, the distribution of $\zeta$ develops a characteristic wavy $m=2$ pattern, in which contours of constant $\zeta$ oscillate in $\varphi$ with large radial amplitude ($1.1\lesssim r\lesssim 1.4$). These oscillations result from passive advection of vortensity by the periodic large amplitude perturbations of $v_r$ associated with the $m=2$ mode. 

Later on, at 425 orbits, one notices two localized vortices (blue dots in Figure \ref{fig:M06.HR.r.a}Ea near $\varphi/\pi\approx 0.2$ and 1.6) appearing quite far from the star, around $r=2.1$. 
These vortices drift radially inwards and eventually merge, resulting in a single vortex visible at 525 orbits at $r\approx 1.9, \varphi/\pi\approx 0.2$, which is responsible for the strong $m=1$ perturbation in $rv_r\sqrt{\Sigma}$ that develops in the outer disk for $r\gtrsim 1.8$. However, careful examination of the vortensity patterns at larger radii reveals that these vortices form early on near the outer boundary of our simulation domain, as a result of a numerical artefact related to our outer boundary condition. 
Their subsequent inward drift is a natural outcome of the vortex dynamics in the disk, see \citet{Paar2010}.

This sequence of vortensity evolution is very typical for our $\M=5$ and 6 runs: we see essentially no vortensity structures produced near the stellar surface (except for the wavy advective patterns), but at late time vortices resulting from numerical artefacts at the outer boundary migrate in and disturb the global vortensity distribution. However, starting at $\M=7$ and higher we never see these numerical artefacts appear in our runs.

\begin{figure}
\ifLOWRES
	\includegraphics[width=\linewidth]{figs/lowres/lowres_M12_spiral.png}
\else
    \includegraphics[width=\linewidth]{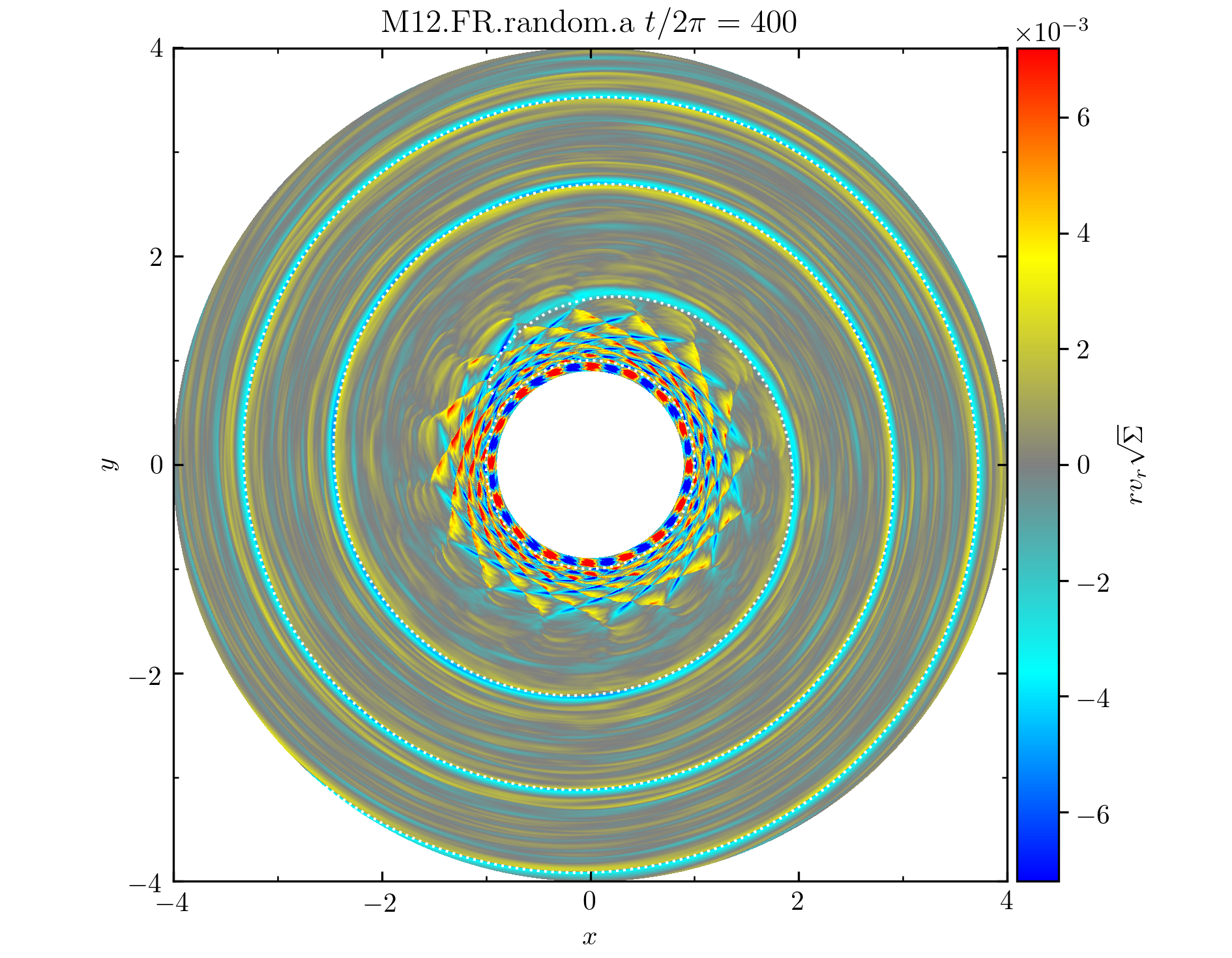}
\fi
    \caption{
    Example of a single-armed spiral emerging in one of our $\M=12$ simulations M12.FR.random.a White dashed curve shows the analytical fit given by equation (\ref{eq:wake}), which matches the shape of the spiral density wave very well. Note a prominent global lower mode active in the inner disk, at $r\lesssim 1.5$. See \S\ref{sect:one-armed} for details.}
    \label{fig:1arm}
\end{figure}


\subsection{Emergence of narrow, single-armed spirals}
\label{sect:one-armed}


In roughly one third of our simulations we observe vortices or rolls to gradually merge into a single strong, coherent vortex, which launches a narrow, single-armed spiral density wave in the disk. A typical example is shown in Fig.~\ref{fig:1arm} illustrating one of our  $\M=12$ runs (M12.FR.random) at 400 orbits. These narrow spiral features form almost exclusively in runs with $\M\ge 7$. This is because, as discussed earlier in \S \ref{sect:vortex=9}-\ref{sect:vortex<9}, single isolated vortices tend to form only in simulations with sufficiently high values of $\M$. These spiral arms are rather long-lived and can last for $\sim 100$ orbits. They are important because they can lead to interesting observational manifestations in the time domain. 

Such single-armed features have much smaller azimuthal width than the $m=1$ patterns emerging in some of the low-$\M$ runs, e.g. the one shown in Figure \ref{fig:M06.HR.r.a}(Fa)-(Fc). They closely resemble the spiral arms that appear in simulations of protoplanetary disks with embedded, moderately massive planets. Because of the narrow azimuthal width, such arms must be superpositions of a number of high-$m$ acoustic modes (as in the case of planet-driven spirals), with pattern speed $\Omega_{\rm p}$ set by the angular frequency $\Omega$ of their parent vortex (or roll).

This allows us to better understand the shape of these arms. Indeed, for $m\gg 1$, the first term in the right-hand side of the WKB dispersion relation (\ref{eq:disp_rel}) can be neglected (at large radii $\Omega(r)$ also becomes small compared to $\Omega_{\rm p}$), allowing us to express
\begin{align}
    k_r(r)\approx - m\frac{\Omega(r)-\Omega_{\rm p}}{c_s(r)},
    \label{eq:k_r}
\end{align}
where we chose sign so that $k_r>0$ in the outer disk, far from the BL. Integrating the relation (\ref{eq:wake_shape}) with this expression for $k_r$ gives the equation for the shape of the wave crest in the form \citep{R02}
\begin{align}
    \varphi(r)=\varphi_{\rm ref}+\int\limits_{r_{\rm ref}}^{r}\frac{\Omega(r^\prime)-\Omega_{\rm p}}{c_s(r^\prime)}dr^\prime,
    \label{eq:wake}
\end{align}
where $\varphi_{\rm ref}$ is the azimuthal coordinate of the wave crest at some reference radius $r_{\rm ref}$. 

The fact that the modes with $m\gg 1$ have $\varphi(r)$ essentially independent of $m$ means that these modes can constructively interfere, maintaining the one-armed profile in a narrow azimuthal range over large radial intervals. For disk-planet interaction this observation was made previously by \citet{OL02} and \citet{R02}, whereas \citet{BZ18} and \citet{MR19} pointed out that this coherence works particularly well in the outer disk (whereas in the inner disk it gets gradually lost). This is relevant for our case since the narrow global arms that we observe are {\it exterior} to the vortices that launch them.

Single-armed spirals that we see in our runs were not observed in previous simulations of the BLs \citep[e.g.][]{BRS12,BRS13a}. Many of these earlier simulations did not extend over the full $2\pi$ in the azimuthal direction, which would both not support single-armed features and affect the emergence and evolution of vortices driving the single-armed spiral. Other simulations that did cover the full $2\pi$ in azimuth had limited radial extent ($r_{\rm max}=2.5R_\star$), which likely prevented them from revealing single-armed spirals. To verify this hypothesis we preformed test runs in which we varied $r_{\rm max}$ and, indeed, did not find any single-armed spirals in simulations with $r_{\rm max}\le 3R_\star$. This suggests that a large radial extent is necessary for capturing the development of such wave phenomena in simulations.


\section{Origin of the vortex-driven modes}
\label{sect:vortex_disc}


In \S \ref{sect:vortex} we uncovered a clear connection between the multiple spiral arms and the vortex-like structures in the near-BL part of the disk. In particular, azimuthal locations of vortices in the near-BL region coincide with the launching sites of the major global spiral arms in the disk. The multiplicity and pattern speeds of these spiral arms are controlled by the number of the corresponding vortices and their radial location. This naturally raises a question of the underlying reason behind this relationship.  

Local perturbations of vortensity, confined both in radius and azimuthal angle, which we call vortices, are known to trigger density waves in accretion disks through the velocity perturbations that they induce in the underlying flow. This has been demonstrated both numerically \citep{Li2001,Johnson2005} and through detailed analytical exploration \citep{Heinemann2009,Paar2010}. In many ways the action of vortices is similar to that of planets (or other massive orbiting perturbers), that launch density waves via their gravitational coupling to the disk at the Lindblad resonances \citep{GT80}. Thus, as long as vortices are present in the inner disk, the excitation of global spiral arms propagating over large distances is quite natural.

However, this brings up the next obvious question: what causes the emergence of vortices in the near-BL region in the first place? We now address this question.

\begin{figure}
	\includegraphics[width=\linewidth]{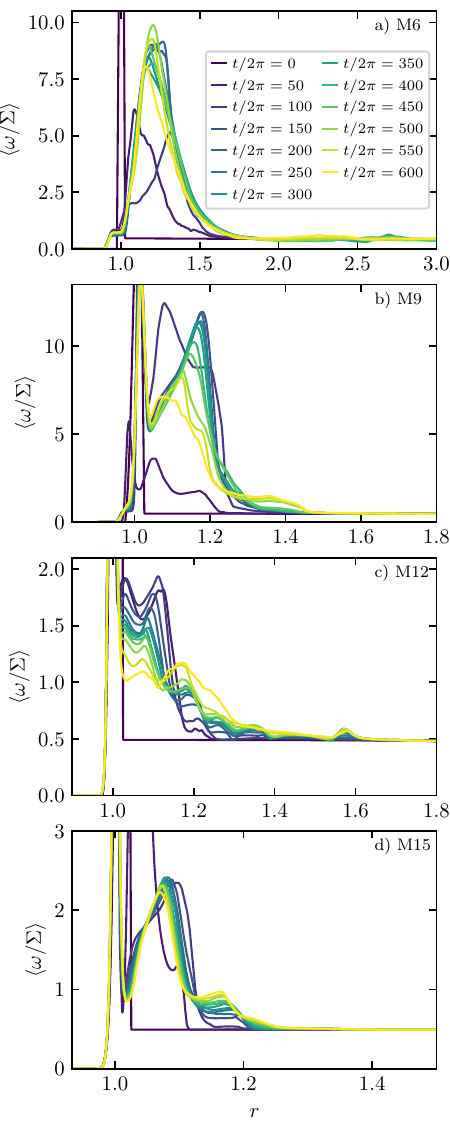}
    \caption{Evolution of the azimuthally-averaged vortensity profile $\zeta(r)=\langle\omega/\Sigma\rangle$ in simulations with different $\M$ (labeled in each panel).  Profiles of $\zeta(r)$ at different times are color-coded according to the legend in the upper panel. See \S\ref{sect:vortex-origin} for the discussion.}
    \label{fig:vort-profiles}
\end{figure}


\subsection{Origin of vortices in the near-BL region}
\label{sect:vortex-origin}


Examination of Figs. \ref{fig:M09.FR.r.a}, \ref{fig:M06.HR.r.a}, \ref{fig:M12.FR.r.a}, \ref{fig:M15.FR.r.a} reveals that in the beginning of the run vortensity grows in the near-BL region above its initial value, which is equal to $\Omega_K/(2\Sigma)$ and is radially constant in the disk for our initial conditions, see \S\ref{sect:ic}. Evolution of $\zeta(r)$ is shown in Fig. \ref{fig:vort-profiles}, where we plot the azimuthally-averaged profiles of the vortensity at different moments of time for the runs described in \S \ref{sect:fiducial}. One can see that in all four runs $\zeta$ experiences substantial evolution in the near-BL region. This raises a possibility of a Rossby wave instability (RWI), which operates in presence of radially structured vortensity, being triggered in this part of the disk. 

The importance of RWI in astrophysical disks has been pointed out by \citet{Lovelace1999} who demonstrated, in particular, that infinitesimal perturbations can grow exponentially provided that the underlying radial profile of $\zeta$ has an extremum. It was subsequently studied by a number of authors both analytically \citep{Li2000,Ono2016} and numerically \citep{Li2001,Johnson2005,Ono2018}. A natural outcome of the nonlinear stage of the RWI is the formation of multiple vortices (each of them launching their individual spiral arms) with their subsequent merger into a single major vortex \citep{Ono2018}. This sequence of events is precisely what we observe in our runs. Previous studies typically triggered the RWI by features in $\zeta(r)$ profile arising due to localized steps, bumps, or gaps in the surface density. The latter --- a drop in $\Sigma$ --- is always found in our simulations, see Figure \ref{fig:M09.FR.r.a_evo_prof}.    

Note that, according to \citet{Pap1989}, in barotropic disks, such as the globally isothermal disk considered in this work, exponentially growing modes of the RWI require the {\it minima} of the vortensity profile to exist in the disk. However, a smooth drop in $\Sigma$ in a Keplerian disk would give rise to a {\it maximum} of $\zeta(r)$, which should be stable according to \citet{Pap1989}. Nevertheless, in many of our simulations we also find the minima of $\zeta(r)$ to emerge quite naturally. Fig. \ref{fig:vort-profiles} shows that at different moments of time $\M=9,12,15$ runs exhibit $\zeta(r)$ profiles with (multiple) local minima, which would give rise to RWI. Interestingly, the profile of $\zeta(r)$ in $\M=6$ run tends to show only a single broad maximum and no minima, see Fig. \ref{fig:vort-profiles}d. This is consistent with the lack of the near-BL vortices in the low-$\M$ runs, see \S \ref{sect:vortex} and Fig. \ref{fig:M06.HR.r.a}. 

At the same time, it should be remembered that derivation of the RWI excitation criterion in \citet{Pap1989} was based on many simplifying assumptions: static, axisymmetric background vortensity profile, infinitesimal perturbations, etc. In real near-BL region, we see that $\zeta(r)$ is generally non-axisymmetric, rapidly changes in time, and is being constantly perturbed by the acoustic waves, which are at least mildly nonlinear. For these reasons the RWI criterion formulated in \citet{Pap1989} may not be directly applicable for interpreting the results of our simulations, even if it seems to work qualitatively. We leave the detailed exploration of the properties of RWI modes in our simulations --- growth rates, pattern speeds, etc. --- to future work.


\subsection{Vortensity evolution in the near-BL region}
\label{sect:vortensity-evolution}


A final step in closing the logical loop of understanding vortex-driven modes is to explain the apparent evolution of vortensity near the stellar surface that we observe in Fig. \ref{fig:vort-profiles}, which is necessary for triggering the RWI. In barotropic disks $\zeta$ is strictly conserved, $\mathrm{d}\zeta/\mathrm{d}t=0$. However, conservation of $\zeta$ gets broken in presence of shocks. In our BL simulations mildly nonlinear  modes evolve into shocks very naturally, driving the growth of vortensity within the resonant cavity where they are trapped. Upper modes do not seem to be efficient at driving the growth of $\zeta$. 

The local rate at which vortensity evolves due to shocks depends on a variety of factors: multiplicity of the waves (i.e. azimuthal wavenumber $m$ of the modes), their amplitude, their pitch angle (depending on the pattern speed $\Omega_{\rm p}$ of the underlying modes), see \citet{Kevlahan1997}, \citet{LinPap2010} and \citet{Dong2011}. In addition to $\zeta$ production at shocks, vortensity is also passively advected into the star as a result of mass accretion. Intricate interplay between these processes leads to a complicated structure in the radial profile of $\zeta(r)$ in the near-BL zone, allowing the RWI to operate.   

To summarize, vortex-driven modes emerge as a result of multi-stage process driven by the sonic modes. First, sonic instability in the supersonic shear layer produces (lower) acoustic modes. Second, these modes, being mildly nonlinear, evolve into shocks and drive vortensity production within the resonant cavity near the stellar surface. Third, accumulation of vortensity creates the conditons for excitation of the RWI, which in turn gives rise to multiple vortices in its nonlinear phase. Finally, each vortex launches a density wave that propagates out from the BL region as a vortex-driven spiral arm. A very similar sequence of steps occurs in tidal coupling of protoplanetary disks with massive embedded planets \citep{Koller2003,Li2005,deVal2007,LinPap2010}: planet-driven density waves shock near the planet, modifying the vortensity profile and triggering RWI, which produces vortices at the edges of the forming gap, with secondary spiral arms being driven by such vortices in the disk.


\subsection{Compact vortices vs ``rolls"}
\label{sect:vortex-vs-roll}


The two main types of vortensity structures that we identify in our simulations --- isolated vortices and rolls --- differ in a number of ways. 

First, rolls tend to appear as azimuthally periodic (often connected) chains of regular patterns of vortensity, whereas isolated vortices have smaller azimuthal extent and are are more irregular in their morphology. Second, isolated vortices are most prominent in the very beginning of the simulation, whereas rolls appear only after sufficient time has passed for the disk surface density and vortensity structure to be substantially modified near the BL. Third, isolated vortices exist only very close to the BL, at $r\to 1$, whereas rolls tend to form at some separation from the BL, typically at $r\sim (1.1-1.5)$.  

At least some of these observations can be interpreted by comparing $\zeta$ maps in Figs. \ref{fig:M09.FR.r.a}, \ref{fig:M06.HR.r.a}, \ref{fig:M12.FR.r.a}, \ref{fig:M15.FR.r.a} with the radial profiles of $\zeta$ in Figure \ref{fig:vort-profiles}. In particular, in the beginning of the simulations $\zeta$ profiles show strong peak of vortensity at $r\to 1$, which is the natural consequence of the initial sharp gradient of $\Omega(r)$ across the BL. These peaks are what gives rise to vortices early on in the simulation. As the run progresses and the BL broadens, radial gradients of $\Omega(r)$ diminish, lowering $\zeta$ peaks at $r\approx 1$ and reducing the significance of the strong, sharp, localized vortices over time.   
Figure \ref{fig:vort-profiles} also shows other vortensity peaks, appearing in the disk at some distance from the BL at later stages. It is easy to see that the radial locations of these peaks coincide with the locations of the chains of rolls that emerge in our runs at roughly the same moments time. In other words, rolls in the inner disk appear to be driven by vortensity generation at weak shocks, into which the near-BL acoustic modes inevitably evolve due to their nonlinear evolution.

Given this difference in origin, one may wonder if isolated vortices are purely an artefact of our initial conditions in the form of a sharp $\Omega(r)$ gradient. This is only partly true, since such gradient persists through our runs because of the $\Omega$ drop in the BL. The amplitude of this gradient (which directly translates into the amplitude of the vortensity peak) is a strong function of $\M$ since the BL width is a sensitive function of the Mach number and scales roughly as $\M^{-2}$, see \citet{BRS12} and Coleman et al. (in prep.). For that reason vortices at $r\to 1$ are never strong in our low-$\M$ runs. This is unlike the high-$\M$ runs, in which the BL is narrow, $\zeta(r)$ maintains a tall peak near the star (see Figure \ref{fig:vort-profiles}) and vortices at $r\approx 1$ tend to be long lived; see e.g. $\zeta$ distribution in the left panels of Figure \ref{fig:M15.FR.r.a}, where some vortex-like structures are present near the BL throughout the full duration of this $\M=15$ run.


\subsection{Robustness of the vortex-driven modes}
\label{sect:vortex-robustness}


Formation of a depression in $\Sigma$ near the stellar surface and the associated peak of $\zeta$ appear essential for providing the conditions for vortex/roll excitation in the near-BL region. Our simulations are inviscid, and such forming gap does not get replenished by the material arriving from larger radii in the disk. However, in real accretion disks there is mass inflow (e.g. due to the MRI), which would tend to refill the gap with gas brought in from larger radii, and might prevent vortex-driven modes from appearing. This possibility may be difficult to realize because of the efficiency with which sonic modes transport mass near the stellar surface. It is plausible that even with the continuous mass inflow from larger radii, sonic modes would still be able to modify $\Sigma(r)$ near the star, sufficient to keep RWI going. And the gap does not need to be very deep for vortex-driven modes to emerge; for example, $\M=12,15$ runs exhibit rather shallow (only $\sim 30\%$ deep) gaps but still support vortex-driven modes. 

Another potential issue with the vortex-driven waves is the fact that our simulations are 2D. While vortices can certainly form in 3D simulations, there is an ongoing debate about their longevity in realistic protoplanetary disks with vertical structure 
\citep{Barranco2005,Lithwick2009,Lesur2009,Meheut2012a,Meheut2012b,Lin2012,Lin2018}. In this regard we note that our own 3D simulations (to be analyzed in the future) do show the emergence of the vortex-driven modes and their survival over long time intervals.


\section{Discussion}
\label{sect:disc}


The main goal of this work is a systematic exploration of the acoustic mode activity in the vicinity of the BL. We do this in a rather simple but easy to control setup, with the Mach number $\M$ being the only key parameter of our runs. The initial distribution of the disk surface density is chosen to ensure a flat vortensity profile, to avoid possible biases related to the initial conditions. 

The equation of state used in this work --- {\it globally} isothermal --- greatly simplifies the analysis of the angular momentum and mass transport in the near-BL region (Coleman et al., in prep), since recent studies \citep{Lin2015,Miranda-ALMA,Miranda-Cooling} have shown that the often used {\it non-barotropic, locally} isothermal equation of state leads to non-conservation of the angular momentum flux carried by the waves even in the absence of explicit dissipation. Our equation of state also allows us to not worry about the long-term effects of heating/cooling on the disk thermal state.

While carrying out this exploration we discovered new types of hydrodynamic wave-like phenomena that emerge in the disk near the stellar surface. Probably the most interesting are the vortex-driven waves, and we already covered their origin and properties at length in \S\S\ref{sect:vortex},\ref{sect:vortex_disc}. In the following we provide a discussion of several other notable results of our simulations, among them the analysis of the regular acoustic (\S\ref{sect:acoustic}) and other (\S\ref{sect:disk-only}) modes, as well as the dependence of their harmonic content on $\M$.


\subsection{Acoustic modes}
\label{sect:acoustic}


Acoustic modes excited by supersonic shear in the BL are interesting not only on their own but also because they are the ultimate drivers of accretion onto the central object (Coleman et al., in prep.) and are intimately involved in generation of other types of modes, see \S\ref{sect:vortex_disc}. Both lower and upper modes (\S\ref{sect:pheno}) are observed in our simulations. Only very rarely we see the third, middle, mode described in \citet{BRS13a} temporarily appear early on in some of our runs. 

We generally find the upper mode to be prominent in the beginning of all our runs. Later on its significance tends to go down in simulations with $\M\lesssim 9$, whereas in simulations with higher $\M$ it may reappear later on, but not always: the upper mode is absent in our $\M=12$ runs but persists through the whole duration of the simulation in $\M=15$ case, see Figures \ref{fig:M12.FR.r.a} \& \ref{fig:M15.FR.r.a}. 

The lower mode is seen in most of our runs, often through the full simulation duration, like in $\M=12$ case (but we remind that $\M=12$ runs are quite unique in maintaining extremely stable lower mode, see \S\ref{sect:M=12}). They are far less prominent in $\M\ge 13$ runs, but are still present there at some level, see below. 

The general expectation following from the theory of acoustic mode excitation outlined in \citet{BR12} and \citet{BRS13a} is that the modes should have comparable strength (in $r v_r\sqrt{\Sigma}$) immediately inside and outside the star. However, very often it is much easier to detect a particular mode inside the star than outside. For example, $\M=6$ run (Figure \ref{fig:M06.HR.r.a}) at $t/2\pi>300$ shows a telltale $k_r=0$ (i.e. radially elongated perturbation pattern) signature of the lower $m=8,9,10,12$ modes inside the star, which do not have a counterpart with matching pattern frequencies outside the star, see Figure \ref{fig:M06.HR.r.a-diag}. We speculate that this departure from the theoretical expectation may be at least partly caused by the non-uniform surface density distribution in the inner disk.

In other cases the apparent lack of the disk counterpart for a mode may be caused by its overlap with some other modes, complicating its identification. This is likely the case for the upper $m=19$ mode with $\Omega_{\rm p}\approx 0.65$ in the $\M=15$ run shown in Figure \ref{fig:M15.FR.r.a}: this mode is obvious inside the star (note its non-zero $k_r$ there), whereas it is hardly visible in the disk. However, Figure \ref{fig:M15.FR.r.a-diag}d shows that this mode is in fact also present in the disk (at $r=1.2$, albeit with a substantially reduced amplitude) with the same $\Omega_{\rm p}$; it is hard to detect by eye in simulation snapshots because of its interference with other modes in the disk. This comparison demonstrates the benefits of automatic mode detection procedure that we employ in analyzing our simulations. 

We now examine how the mix of modes detected by our automated analysis compares with the dispersion relations derived in \citet{BRS13a} and this work, see \S\ref{sect:upper},\ref{sect:lower}. 


\subsubsection{Dispersion relation for acoustic modes}
\label{sect:acoustic-DR}


Figure \ref{fig:multi_dispersion} displays the $(m,\Omega_{\rm p})$ pairs for the modes found by our automatic mode detection procedure in runs with different values of $\M$. Some of these modes are truly global (green stars), i.e. they are detected as a wave pattern with the same $m$ and $\Omega_{\rm p}$ {\it at the same interval of time} both inside the star and in the disk (at $r=1.2$). In most cases modes are found only in the disk (yellow pluses) or only in the star (blue circles), a possibility that we mentioned earlier.

We also display in red the dispersion relations (\ref{eq:lower-DR}) for the lower modes (dot-dashed) and (\ref{eqn:upper_disp1}) for the upper modes (dotted). Note that equation (\ref{eq:lower-DR}) depends on a parameter $r_0$ (specific for each $\M$), which we fix by aligning the lower mode dispersion relation curves with the clusters of $(m,\Omega_{\rm p})$ points in Figure \ref{fig:multi_dispersion}; this procedure is not very straightforward for $\M=14,15$, see panels (k) and (l). We also note that the dispersion relation (\ref{eqn:upper_disp1}) assumes that a plateau in $\Omega(r)$ has already developed near the stellar surface (see Figure \ref{fig:M09.FR.r.a_evo_prof}a), so that the epicyclic frequency is $\kappa\approx 2\Omega$; this may not be true early on in the simulation. The ``height" of this plateau at late times $\Omega_{\rm max}$, i.e. the maximum value of $\Omega$, is shown by the horizontal dashed curves.  

In general we see good correspondence between the dispersion relations and the detected modes, as typically a significant number of $(m,\Omega_{\rm p})$ points fall on top of the red curves. These modes often cover a significant range of azimuthal wavenumbers (e.g. the lower modes in panels (f) or (h)), and sometimes come in clusters, i.e. are grouped in $m$ (e.g. $m=10-16$ in panel (e), or $m=21-26$ in panel (g)). Such groupings likely result from the nonlinear evolution of a single dominant mode: nonlinear distortion of a perturbation profile, natural for even moderately strong acoustic waves that we see in our runs, transfers power to other modes, primarily to those with similar $m$. This likely explains the slow but persistent changes in the modes that we observe: as the acoustic waves are dispersive, the new modes produced as a result of the non-linear evolution of a parent mode eventually lose coherence with it, smearing out the original wave packet. Thus, the finite lifetime of the modes that we see in our runs should not come as a surprise.  

At the same time, there are some modes that do not line up with the red curves. Many of them are simply not the usual upper and lower acoustic modes, see e.g. \S\ref{sect:disk-only}. But many others end up being the higher-order azimuthal harmonics of the primary modes. To illustrate that we show the harmonics of the main dispersion relation branches with twice and three times higher azimuthal wavenumber $m$ and the same $\Omega_{\rm p}$ as black dotted and dot-dashed curves in Figure \ref{fig:multi_dispersion}. One can see that in some cases the modes lying on the main branch of the dispersion relation have counterparts with the same $\Omega_{\rm p}$ at or close to one of the higher-order branches of that dispersion relation. Clear examples of this can be seen in panel (b) for the lower modes with $\Omega_{\rm p}\in (0.45,0.55)$, and in panel (g) for the upper modes with $\Omega_{\rm p}\in (0.5,0.6)$. Such higher-order azimuthal counterparts of the modes naturally result from the {\it non-sinusoidal shape} of the wave packets with certain azimuthal periodicity. 

For almost all values of $\M$ we also see some modes that stay close to the $\Omega_{\rm p}=\Omega_{\rm max}$ horizontal dashed line. These modes must be trapped in the innermost part of the disk where at late times $\Omega(r)$ features a plateau with $\Omega(r)\approx \Omega_{\rm max}$. They are likely related to the vortensity structures forming in this part of the disk --- vortices or rolls, which are passively advected with the fluid at almost constant orbital frequency $\Omega_{\rm max}$. Stability of $\Omega_{\rm max}$ (see Figure \ref{fig:M09.FR.r.a_evo_prof}) should help these modes maintain their coherence over long intervals of time, which may have important implications for the variability associated with the BL (such as dwarf nova oscillations); this issue will be explored in a future work.  Also, for $\M<8$ some lower modes feature $\Omega_{\rm p}$ exceeding $\Omega_{\rm max}$; these modes must have been present in the disk early on, when the $\Omega(r)$ profile was still close to Keplerian. 

Finally, panels (e) and (f) of Figure \ref{fig:multi_dispersion} compare two simulations with the same Mach number $\M=9$ but different resolutions. The higher resolution run M09.HR.r.a ($8192\times 8192$) appears to show no disk modes, in contrast with the run M09.FR.r.a ($4096\times 4096$), which reveals a number of both global and disk-only modes. However, this outcome is caused simply by the difficulty of mode detection by our automated mode-finding algorithm in the higher resolution run: by examining its outputs by eye we do find a number of disk modes, which simply fluctuate a bit more than is allowed by our software to register them as waves with a well-defined $\Omega_{\rm p}$ (see Appendix \ref{sec:mode_detect} for details).


\subsection{Other disk-only modes}
\label{sect:disk-only}


Next we briefly discuss a couple of other wave structures that are seen in our simulations and cannot be classified as upper or lower modes (or their harmonics). These modes are present only in the disk, with no counterpart inside the star.


\subsubsection{Resonant modes}
\label{sect:low-m_disc}

One type of such waves are the relatively low-$m$ modes in the disk trapped in the resonant cavity near the star; for this reason these waves may be confused with the usual lower modes. However, unlike the lower modes they (1) do not have a strong counterpart with the same $m$ inside the star, (2) usually do not exhibit densely criss-crossed pattern, and (3) obey a very different dispersion relation, as we show next. The difference in appearance between the two types of modes can be easily spotted in Figure \ref{fig:M12.FR.r.a} for $\M=12$, where a strong lower mode is present in panels with $t/2\pi=150-450$, whereas at $t/2\pi=550$ there is a strong $m=7$ disk-only mode with no crossings of the incoming and outgoing wakes, trapped at $r<1.25$. 

\begin{figure}
	\includegraphics[width=1.0\linewidth]{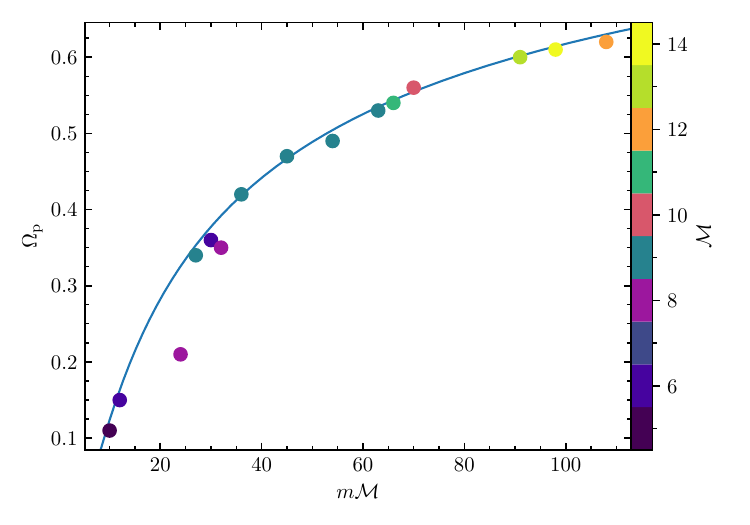}
    \caption{Dispersion relation (\ref{eq:disk-mode-app}) for the resonant modes for $q=2$, plotted together with the $(m\M,\Omega_{\rm p})$ values for the resonant modes identified in our simulations, shown with points colored by their value of $\M$. The bright magenta points correspond to the $\M=8$ simulation, which show the largest deviation from the trend. See \S\ref{sect:low-m_disc} for details.
    }
    \label{fig:res-modes}
\end{figure}

This type of mode manifests itself also at other values of $\M$: as a strong $m=2$ mode for $\M=6$ (at $t/2\pi=325-425$), confined to $r<2.5$; as a $m=6$ mode for $\M=9$ (at $t/2\pi=400-450$), confined to $r<1.3$ (although strongly disturbed by the $m=5$ vortex-driven mode); and a $m\approx 20$ mode at $\M=15$ (at $t/2\pi=275-5756$), confined to $r<1.17$. We also find this mode to persist in our very long $\M=9$ run, where it dominates as either $m=3$ or $m=4$ pattern for more than 2000 orbits.

Such modes were previously described in \citet{BRS12}, who traced their origin to a {\it geometric resonance} for a trapped acoustic wave: if, after multiple reflections off the stellar surface and the Inner Lindblad Resonance, the density wave closes on itself (after its azimuthal phase wraps around the star $q$ times, where $q$ is a small integer), then this reinforces its strength and gives rise to a stable mode. Mathematically, \citet{BRS12} have shown that this leads to a relationship between $m$ and $\Omega_{\rm p}$ for these modes, which can be cast as
\ba  
q\frac{\pi}{\M}=\frac{1}{\Omega(R_\star)R_\star}\int_{R_\star}^{r_{\rm ILR}}
dr\sqrt{m^2\left[\Omega(r)-\Omega_{\rm p}\right]^2-\kappa^2(r)},
\label{eq:disk-mode}
\ea  
where $r_{\rm ILR}$ is given by equation (\ref{eq:r_ILR}). It was also shown in that work that such {\it resonant} modes indeed obey the dispersion relation (\ref{eq:disk-mode}), see their Fig. 18 in \citet{BRS12}.

One could turn the integral relationship between $m$ and $\Omega_{\rm p}$ in equation (\ref{eq:disk-mode}) into an approximate algebraic one by dropping the $\kappa^2$ term; this is equivalent to approximating $r_{\rm ILR}\approx r_c$ and is accurate for $m\gg 1$, see equation (\ref{eq:r_ILR}) and Figure \ref{fig:sonic-examples}b. Then one finds
\ba  
m\M\approx q\pi\left[2+\frac{\Omega_{\rm p}}{\Omega_K(R_\star)}-3\left(\frac{\Omega_{\rm p}}{\Omega_K(R_\star)}\right)^{1/3}\right]^{-1}.
\label{eq:disk-mode-app}
\ea  
As $\M$ in our runs increases, we find the resonant mode wavenumber $m$ to increase as well. As a result, equation (\ref{eq:disk-mode-app}) predicts that $\Omega_{\rm p}$ for this mode should also {\it increase} with both its $m$ and $\M$, see Figs. 17 and 18 of \citet{BRS12}. This leads to narrowing of the resonant cavity for this mode as $\M$ goes up, just as we find in our runs. 

One can see that $m$ and $\M$ enter equation (\ref{eq:disk-mode-app}) only in combination $m\M$. This allows us to plot the dispersion relation (\ref{eq:disk-mode-app}) as a single curve in $(m\M,\Omega_{\rm p})$ coordinates for runs with different $\M$. We do this in Figure \ref{fig:res-modes}, where we also plot $(m\M,\Omega_{\rm p})$ points for all resonant modes that we were able to reliably identify in our simulations. One can see that with $q=2$ the dispersion relation (\ref{eq:disk-mode-app}) fits the simulation results quite well. The only exception are the two occurrences of the resonant mode in our $\M=8$ run, for which a different value of $q$ might have worked better as we see multiple crossings of resonant modes in this run (usually we see only a single crossing). Note that $q=2$ that we find in this work is different from $q=1$ found in \citet{BRS12}, not clear why. 

The dispersion relation shown in Figure \ref{fig:res-modes} is clearly different from that of the lower modes, for which $\Omega_{\rm p}$ always {\it decreases} with $m$, see Figure \ref{fig:multi_dispersion}. This is despite the fact that the two types of modes have similar morphological appearance and are confined to a resonant cavity in the disk; they also have a similar effect on the angular momentum and mass transport in the disk (Coleman et al., in prep.). 

At the same time, the dispersion relation (\ref{eq:DR-uppr-approx}) for the upper acoustic mode (as well as its more refined version (\ref{eqn:upper_disp1})) leads to $\Omega_{\rm p}$ {\it increasing} with $m$, similar to the behavior predicted by the equation (\ref{eq:disk-mode-app}). For that reason, in Figure \ref{fig:multi_dispersion} we often find the $(m,\Omega_{\rm p})$ pairs for the resonant modes to lie close to the main branch of the upper mode dispersion relation, e.g. see $m=2$, $\Omega_{\rm p}\approx 0.15$ resonant mode for $\M=6$ in Figure \ref{fig:multi_dispersion}b, or $m=6$, $\Omega_{\rm p}\approx 0.5$ resonant mode for $\M=9$ in Figure \ref{fig:multi_dispersion}e.


\subsubsection{Low-$m$, $k_r=0$ modes}
\label{sect:krzero_disc}

As mentioned in \S\ref{sect:fiducialM=9}, our fiducial $\M=9$ run shows yet another disk-only mode with $m=2$, $\Omega_{\rm p}\approx 0.15$, between roughly 150 and 400 orbits in Figure \ref{fig:M09.FR.r.a}. It has a very unusual appearance, with $k_r=0$ and azimuthally extended perturbation pattern (i.e. not a narrow feature), confined to $r\lesssim 2.2$, which is close to the $r_{\rm ILR}$ for this mode. Its perturbation also undergoes a flip by $\pi$ in azimuthal phase at $r\approx 1.5$. Vortensity maps in Figure \ref{fig:M09.FR.r.a} show no structures in $\zeta$ at this radius or beyond it.

The emergence of this mode is not unique to the run M09.FR.r.a displayed in Figure \ref{fig:M09.FR.r.a}, as we observe it in several other $\M=9$ runs with different kinds of initial perturbation. The low $m$ and $\Omega_{\rm p}$ of this mode places it very close to the main branch of the upper mode dispersion relation (\ref{eqn:upper_disp1}), see Figure \ref{fig:multi_dispersion}e. This is not surprising since that dispersion relation was derived assuming $k_r=0$ (see \S\ref{sect:upper-DR}), which is true at all $r$ for the $m=2$ mode that we see. At the moment we do not have an explanation for the origin or properties of this unusual disk-only mode. 


\subsection{Dominant modes as a function of $\M$}
\label{sect:dom_modes}


Given that we have BL simulations for every integer value of $\M$ between 5 and 15, we can explore how the azimuthal periodicity of the modes that we detect changes with $\M$. In general, we find that a particular mix of modes that exist at different times in a given simulation is pretty stochastic. This means that a different realization of the same simulation, especially with the different model of initial noise introduced to trigger the instability (\S\ref{sect:ic}), would result in a somewhat different outcome in terms of the mode types and azimuthal wavenumbers $m$. The only notable exception are our $\M=12$ simulations, in all of which we robustly see the $m=16$ lower mode with a pattern speed $\Omega_{\rm p}=0.45$ dominating both inside and outside of the star over hundreds of orbits. Resolution of the simulations also plays a role, see Figure \ref{fig:multi_dispersion}e,f, but the differences there often depend on the performance of our mode-finding analysis software, see \S\ref{sect:acoustic-DR}.


\begin{figure}
	\includegraphics[width=\linewidth]{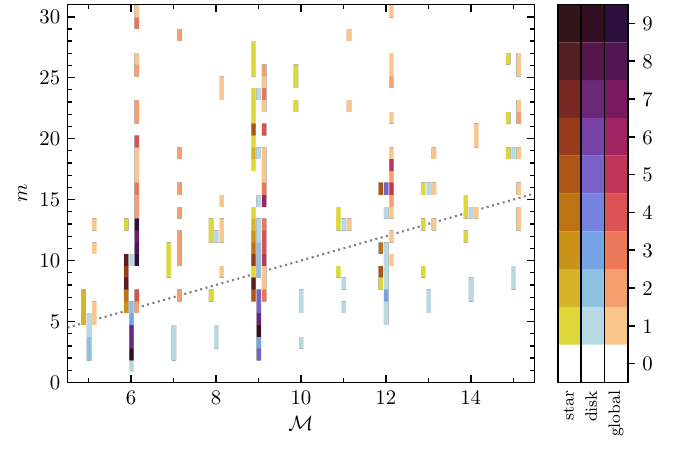}
    \caption{Histogram of the azimuthal wave number $m$ for the dominant modes identified in simulations with different values of Mach number $\mathcal{M}$. For each $\mathcal{M}$ there are three histograms with arbitrary horizontal displacement. Left (yellow-orange): dominant star-only modes. Middle (blue-indigo): dominant disk-only modes. Right (pink-purple): global modes. For each $\M$ the dominant modes are defined as the three modes with the highest time-integrated power, volume averaged over a specific region. For some values of $\mathcal{M}$ we have multiple simulations, so the numeric value of the corresponding bin reflects the number of runs, in which the mode meets the aforementioned criteria. The dotted line corresponds to $\mathcal{M}=m$ which qualitatively follows the trend seen in the data. See \S\ref{sect:dom_modes} for details.}
    \label{fig:mode_hist}
\end{figure}


On the other hand, we do observe certain trends with $\M$. In particular, Figure \ref{fig:multi_dispersion} reveals that the lower modes --- $(m,\Omega_{\rm p})$ points aligned with the lower mode branches --- are more common for $\M\le 12$, whereas the upper modes start showing up in noticeable clusters along the upper dispersion relation branches for $\M\ge 9$. 

To examine possible trends with $\M$ at a more quantitative level, we carried out the following exercise. First, we compute the power in all modes with $m<32$ for the variable $v_r\sqrt{\Sigma}$ and then integrate it over time for $t/2\pi>100$ and over radius in three distinct domains: ``star" defined as $r<1$, ``disk"  defined as $1<r<2.2$, and ``global" defined as $r<2.2$, i.e. ``disk+star". In a given domain, the three modes with the highest time- and radius- integrated power are considered to be the dominant mode. This data is summarized in Fig.~\ref{fig:mode_hist}, where the histograms of different color characterize the distribution of $m$ for the dominant modes in three respective regions for all $\M$.

By examining this figure we find that at each $\mathcal{M}$ there is a substantial spread in the values of $m$, even in a given domain. This spread is caused by a number of factors: stochasticity of the mix of modes, different types of modes involved (e.g. upper, lower, disk-only), resolution, etc. Also, we have reasonably representative statistics on the distribution of $m$ only for $\M=6,9,12$, for which there are multiple runs with different initial conditions and resolutions; for most other values of $\M$ only a single run is available. 


Qualitatively, there is an overall trend of increasing the dominant mode wave number $m$ with growing Mach number $\mathcal{M}$. Just as a guide, dotted line in Figure \ref{fig:mode_hist} shows a linear relation $m=\M$. This line does not represent a fit of any kind and is merely shown to guide the eye. One conclusion that we can draw from this exercise is that a complete characterization of the mix of the dominant modes operating in the vicinity of the BL may require a substantially larger number of simulations than we have presented in this work.


\subsection{Sensitivity to numerical parameters}
\label{sect:num_convergence}


Our simulation suite allows us to probe the sensitivity of the results to certain numerical inputs for some values of $\M$, namely the initial noise pattern used to trigger the sonic instability in the BL (\S\ref{sect:ic}) and resolution (\S\ref{sect:conv}), see Table \ref{table:allruns}  (sensitivity to boundary conditions has been already discussed in \S\ref{sect:conv},\ref{sect:vortex<9},\ref{sect:one-armed}). 

When comparing the simulations with the different forms of the initial noise (run at the same resolution and $\M$), we generally do not find strong differences or trends with the noise pattern. For $\M=6,9$ the qualitative behavior of the simulations remains the same, although, as we alluded to in \S\ref{sect:other-M9},\ref{sect:dom_modes}, the detailed outcomes of individual simulations are stochastic. And all $\M=12$ runs are similar to one another even at the quantitative level, see \S\ref{sect:M=12}.

Resolution has a more substantial effect on our results. For $\M=9$ it affects azimuthal wavenumber $m$ of the dominant modes, with $m$ increasing with resolution. For example, in simulations with the same block random initial condition ``r.a", we find that the dominant lower mode has $m=7$ at lowest resolution ($2048\times 2048$), increasing to $m=19$ in the fiducial resolution ($4096\times 4096$) case, and reaches $m=26$ at the highest resolution ($8192\times 8192$). The transitions between the different types of modes described in \S\ref{sect:fiducialM=9} are captured quite reliably between the high and fiducial resolution cases, suggesting that their results are converged at least at the qualitative level, but less so at the lowest resolution.

The dependence on resolution is stronger in the $\M=6$ simulations. In particular, high resolution ($2048\times 2048$) $\M=6$ runs demonstrate the early development of the low-$m$ ($m=2,3$) resonant modes, typically around $t/2\pi=200$, while the low resolution ($1024\times 1024$) runs either do not show these modes at all, or exhibit them very late. Thus, high resolution is clearly necessary for revealing important features of the BLs with low $\M$.


\subsection{Comparison with the existing studies}
\label{sect:literature}


A number of past studies of the BLs, both (semi-)analytical \citep{Kip1978,NAR93,POP93} and numerical \citep{Kley1996,ARM02,Stein2002,BAL09,Hert2017}, postulate some form of {\it local shear} stress to enable angular momentum transport inside the BL. Since in practice transport in the BL is mediated by the global acoustic modes \citep{BRS12,BRS13a}, these studies cannot be directly compared to our work.

Our study goes beyond (in ways already discussed in \S\ref{sect:sim}) the similar past work of \citet{BRS12,BRS13a,BRS13b} and \citet{Hert2015}, who also simulated BLs mediated by the acoustic waves. We explore a larger set of Mach number values, use higher resolution and longer run times, and  carry out an extensive exploration of the sensitivity of our results to resolution and initial conditions. We also provide a very careful analysis of our results and extensively study the harmonic content of our simulations. All this led to new important findings such as the vortex-driven modes (\S\ref{sect:vortex}), one-armed spirals (\S\ref{sect:one-armed}), and so on.

\citet{BEL17} considered a different way of exciting acoustic modes in the disk, namely by coupling them to the incompressible inertial waves inside the star. Our use of the globally isothermal equation of state precludes us from exploring this possibility, which should be addressed by future simulations with more sophisticated treatment of gas thermodynamics. 

Finally, in their 3D, unstratified MHD simulations \citet{Bel2018} found that acoustic waves do not efficiently remove angular momentum from the accreting gas in the BL, causing a dense belt of rapidly spinning material to form near the stellar equator. While this is an important issue, which should be addressed in the future using stratified MHD simulations with realistic thermodynamics, \citet{Bel2018} do find acoustic waves to be active in the disk, which is what our study focused on.


\subsection{Observational implications}
\label{sect:observations}


Observational implications of the wave-driven angular momentum transport in the BL have been previously discussed in \citet{BRS12,BRS13a}. One of them is the modification of the spectral signature associated with the energy dissipation in the BL. While the energy conservation implies that the total amount of energy released by the accreted matter must be large, a particular band in which the associated emission is released should be dramatically affected by the global nature of the angular momentum and energy transport by the acoustic modes. This is likely to have important ramifications for the so-called ``missing boundary layer" problem \citep{Ferland1982}. 

Long-lived mode patterns that we see in our simulations should also lead to characteristic variability associated with them. Our current work opens up new possibilities in this regard, by revealing the existence of the vortex-driven modes and one-armed spiral arms in the disk. Both of them may affect the light curves of objects accreting through the BL in characteristic ways. These (and other) implications of the wave-mediated accretion through the BL will be further investigated in the upcoming work.


\section{Summary}
\label{sect:sum}


In this paper, first in a series, we presented a suite of global, 2D, hydrodynamic simulations of the BLs using simple thermodynamics (globally isothermal) and encompassing both the outermost layers of the accreting object and a substantial region of the inner disk. Using this data set we carried out a systematic exploration of the different waves that emerge in the disk in the vicinity of the BL (and regulate its properties), as a function of Mach number of the system $\M$. Our key results 
can be summarized as follows.

\begin{itemize}

\item We discover a new class of modes that are triggered by the emergence of non-axisymmetric, localized vortensity structures in the vicinity of the BL. These vortex-driven modes are quite prominent in many of our simulations. We argue that their parent vortices result from the Rossby wave instibility, triggered by the vortensity production near the BL driven by the nonlinear damping of acoustic modes. 

\item In many simulations we observe multiple near-BL vortices to merge into a single one, giving rise to a prominent global, one-armed spiral density wave in the disk. Such structures may naturally cause periodicity of the BL emission. 

\end{itemize}

We can also make the following statements about the wave patterns emerging in our runs.

\begin{itemize}

\item Different types of modes can easily co-exist in the inner disk. They have finite life times, although some of them can operate for hundreds of orbits. While some of the modes that we see are global, i.e. operate both in the disk and the star, others are star-only or disk-only modes. 

\item We compared characteristics of many different modes identified in our simulations with their analytical dispersion relations and found good agreement. 

\item With rare exceptions, a particular mix of modes that we find in our runs (with slightly different initial conditions) is somewhat stochastic. As the Mach number of our simulations changes we find the mix of near-BL modes to evolve, with the azimuthal wavenumber of the dominant modes showing tendency to increase with $\M$. 

\end{itemize}

Our results pave the way for future efforts to explore angular momentum and mass transport, as well as the associated evolution of the disk in the vicinity of the BL, driven by the modes identified in this study.

\section*{Acknowledgements}

We thank Changgoo Kim for his assistance in modifying the FFT module in \athena and Jim Stone for making the code publicly available. We gratefully acknowledge financial support from NSF via grant AST-1515763, NASA via grant 14-ATP14-0059, and Institute for Advanced Study via the John N. Bahcall Fellowship to R.R.R. Research at the Flatiron Institute is supported by the Simons Foundation.
Resources supporting this work were provided by the NASA High-End Computing (HEC) Program through the NASA Advanced Supercomputing (NAS) Division at Ames Research Center.
Through allocation AST160008, this work used the Extreme Science and Engineering Discovery Environment (XSEDE), which is supported by National Science Foundation grant number ACI-1548562 \citep{XSEDE}.

\section*{Data Availability}

The data underlying this article will be shared on reasonable request to the corresponding author.




\bibliographystyle{mnras}
\bibliography{citations}

\begin{thebibliography}{}
\makeatletter
\relax
\def\mn@urlcharsother{\let\do\@makeother \do\$\do\&\do\#\do\^\do\_\do\%\do\~}
\def\mn@doi{\begingroup\mn@urlcharsother \@ifnextchar [ {\mn@doi@}
  {\mn@doi@[]}}
\def\mn@doi@[#1]#2{\def\@tempa{#1}\ifx\@tempa\@empty \href
  {http://dx.doi.org/#2} {doi:#2}\else \href {http://dx.doi.org/#2} {#1}\fi
  \endgroup}
\def\mn@eprint#1#2{\mn@eprint@#1:#2::\@nil}
\def\mn@eprint@arXiv#1{\href {http://arxiv.org/abs/#1} {{\tt arXiv:#1}}}
\def\mn@eprint@dblp#1{\href {http://dblp.uni-trier.de/rec/bibtex/#1.xml}
  {dblp:#1}}
\def\mn@eprint@#1:#2:#3:#4\@nil{\def\@tempa {#1}\def\@tempb {#2}\def\@tempc
  {#3}\ifx \@tempc \@empty \let \@tempc \@tempb \let \@tempb \@tempa \fi \ifx
  \@tempb \@empty \def\@tempb {arXiv}\fi \@ifundefined
  {mn@eprint@\@tempb}{\@tempb:\@tempc}{\expandafter \expandafter \csname
  mn@eprint@\@tempb\endcsname \expandafter{\@tempc}}}

\bibitem[\protect\citeauthoryear{{Armitage}}{{Armitage}}{2002}]{ARM02}
{Armitage} P.~J.,  2002, \mn@doi [\mnras] {10.1046/j.1365-8711.2002.05152.x},
  \href {http://adsabs.harvard.edu/abs/2002MNRAS.330..895A} {330, 895}

\bibitem[\protect\citeauthoryear{{Bae} \& {Zhu}}{{Bae} \& {Zhu}}{2018}]{BZ18}
{Bae} J.,  {Zhu} Z.,  2018, \mn@doi [\apj] {10.3847/1538-4357/aabf8c}, \href
  {https://ui.adsabs.harvard.edu/abs/2018ApJ...859..118B} {859, 118}

\bibitem[\protect\citeauthoryear{{Balbus} \& {Hawley}}{{Balbus} \&
  {Hawley}}{1991}]{BAL91}
{Balbus} S.~A.,  {Hawley} J.~F.,  1991, \mn@doi [\apj] {10.1086/170270}, \href
  {http://adsabs.harvard.edu/abs/1991ApJ...376..214B} {376, 214}

\bibitem[\protect\citeauthoryear{{Balsara}, {Fisker}, {Godon}  \&
  {Sion}}{{Balsara} et~al.}{2009}]{BAL09}
{Balsara} D.~S.,  {Fisker} J.~L.,  {Godon} P.,   {Sion} E.~M.,  2009, \mn@doi
  [\apj] {10.1088/0004-637X/702/2/1536}, \href
  {http://adsabs.harvard.edu/abs/2009ApJ...702.1536B} {702, 1536}

\bibitem[\protect\citeauthoryear{{Barranco} \& {Marcus}}{{Barranco} \&
  {Marcus}}{2005}]{Barranco2005}
{Barranco} J.~A.,  {Marcus} P.~S.,  2005, \mn@doi [\apj] {10.1086/428639},
  \href {https://ui.adsabs.harvard.edu/abs/2005ApJ...623.1157B} {623, 1157}

\bibitem[\protect\citeauthoryear{{Belyaev}}{{Belyaev}}{2017}]{BEL17}
{Belyaev} M.~A.,  2017, \mn@doi [\apj] {10.3847/1538-4357/835/2/238}, \href
  {http://adsabs.harvard.edu/abs/2017ApJ...835..238B} {835, 238}

\bibitem[\protect\citeauthoryear{{Belyaev} \& {Quataert}}{{Belyaev} \&
  {Quataert}}{2017}]{BQ17}
{Belyaev} M.~A.,  {Quataert} E.,  2017, preprint, \href
  {http://adsabs.harvard.edu/abs/2017arXiv170901197B} {} (\mn@eprint {arXiv}
  {1709.01197})

\bibitem[\protect\citeauthoryear{{Belyaev} \& {Quataert}}{{Belyaev} \&
  {Quataert}}{2018}]{Bel2018}
{Belyaev} M.~A.,  {Quataert} E.,  2018, \mn@doi [\mnras]
  {10.1093/mnras/sty860}, \href
  {https://ui.adsabs.harvard.edu/abs/2018MNRAS.479.1528B} {479, 1528}

\bibitem[\protect\citeauthoryear{{Belyaev} \& {Rafikov}}{{Belyaev} \&
  {Rafikov}}{2012}]{BR12}
{Belyaev} M.~A.,  {Rafikov} R.~R.,  2012, \mn@doi [\apj]
  {10.1088/0004-637X/752/2/115}, \href
  {http://adsabs.harvard.edu/abs/2012ApJ...752..115B} {752, 115}

\bibitem[\protect\citeauthoryear{{Belyaev}, {Rafikov}  \& {Stone}}{{Belyaev}
  et~al.}{2012}]{BRS12}
{Belyaev} M.~A.,  {Rafikov} R.~R.,   {Stone} J.~M.,  2012, \mn@doi [\apj]
  {10.1088/0004-637X/760/1/22}, \href
  {http://adsabs.harvard.edu/abs/2012ApJ...760...22B} {760, 22 (BRS)}

\bibitem[\protect\citeauthoryear{{Belyaev}, {Rafikov}  \& {Stone}}{{Belyaev}
  et~al.}{2013a}]{BRS13a}
{Belyaev} M.~A.,  {Rafikov} R.~R.,   {Stone} J.~M.,  2013a, \mn@doi [\apj]
  {10.1088/0004-637X/770/1/67}, \href
  {http://adsabs.harvard.edu/abs/2013ApJ...770...67B} {770, 67 (BRS1)}

\bibitem[\protect\citeauthoryear{{Belyaev}, {Rafikov}  \& {Stone}}{{Belyaev}
  et~al.}{2013b}]{BRS13b}
{Belyaev} M.~A.,  {Rafikov} R.~R.,   {Stone} J.~M.,  2013b, \mn@doi [\apj]
  {10.1088/0004-637X/770/1/68}, \href
  {http://adsabs.harvard.edu/abs/2013ApJ...770...68B} {770, 68 (BRS2)}

\bibitem[\protect\citeauthoryear{{Binney} \& {Tremaine}}{{Binney} \&
  {Tremaine}}{2008}]{BT}
{Binney} J.,  {Tremaine} S.,  2008, {Galactic Dynamics: Second Edition}

\bibitem[\protect\citeauthoryear{{Chandrasekhar}}{{Chandrasekhar}}{1960}]{Chandra}
{Chandrasekhar} S.,  1960, \mn@doi [Proceedings of the National Academy of
  Science] {10.1073/pnas.46.2.253}, \href
  {https://ui.adsabs.harvard.edu/abs/1960PNAS...46..253C} {46, 253}

\bibitem[\protect\citeauthoryear{{Coleman}, {Kotko}, {Blaes}, {Lasota}  \&
  {Hirose}}{{Coleman} et~al.}{2016}]{COL16}
{Coleman} M.~S.~B.,  {Kotko} I.,  {Blaes} O.,  {Lasota} J.-P.,   {Hirose} S.,
  2016, \mn@doi [\mnras] {10.1093/mnras/stw1908}, \href
  {http://adsabs.harvard.edu/abs/2016MNRAS.462.3710C} {462, 3710}

\bibitem[\protect\citeauthoryear{{Coleman}, {Blaes}, {Hirose}  \&
  {Hauschildt}}{{Coleman} et~al.}{2018}]{COL18}
{Coleman} M. S.~B.,  {Blaes} O.,  {Hirose} S.,   {Hauschildt} P.~H.,  2018,
  \mn@doi [\apj] {10.3847/1538-4357/aab6a7}, \href
  {https://ui.adsabs.harvard.edu/#abs/2018ApJ...857...52C} {857}

\bibitem[\protect\citeauthoryear{{Dong}, {Rafikov}  \& {Stone}}{{Dong}
  et~al.}{2011}]{Dong2011}
{Dong} R.,  {Rafikov} R.~R.,   {Stone} J.~M.,  2011, \mn@doi [\apj]
  {10.1088/0004-637X/741/1/57}, \href
  {https://ui.adsabs.harvard.edu/abs/2011ApJ...741...57D} {741, 57}

\bibitem[\protect\citeauthoryear{{Drury}}{{Drury}}{1979}]{Drury1979}
{Drury} L.~O.,  1979, PhD thesis, University of Cambridge, UK

\bibitem[\protect\citeauthoryear{{Drury}}{{Drury}}{1980}]{Drury1980}
{Drury} L.~O.,  1980, \mn@doi [\mnras] {10.1093/mnras/193.2.337}, \href
  {https://ui.adsabs.harvard.edu/abs/1980MNRAS.193..337D} {193, 337}

\bibitem[\protect\citeauthoryear{{Drury}}{{Drury}}{1985}]{Drury1985}
{Drury} L.~O.,  1985, \mn@doi [\mnras] {10.1093/mnras/217.4.821}, \href
  {https://ui.adsabs.harvard.edu/abs/1985MNRAS.217..821D} {217, 821}

\bibitem[\protect\citeauthoryear{{Ferland}, {Langer}, {MacDonald}, {Pepper},
  {Shaviv}  \& {Truran}}{{Ferland} et~al.}{1982}]{Ferland1982}
{Ferland} G.~J.,  {Langer} S.~H.,  {MacDonald} J.,  {Pepper} G.~H.,  {Shaviv}
  G.,   {Truran} J.~W.,  1982, \mn@doi [\apjl] {10.1086/183910}, \href
  {https://ui.adsabs.harvard.edu/abs/1982ApJ...262L..53F} {262, L53}

\bibitem[\protect\citeauthoryear{{Ghosh}, {Lamb}  \& {Pethick}}{{Ghosh}
  et~al.}{1977}]{Ghosh1977}
{Ghosh} P.,  {Lamb} F.~K.,   {Pethick} C.~J.,  1977, \mn@doi [\apj]
  {10.1086/155606}, \href
  {https://ui.adsabs.harvard.edu/abs/1977ApJ...217..578G} {217, 578}

\bibitem[\protect\citeauthoryear{{Gilfanov}, {Revnivtsev}  \&
  {Molkov}}{{Gilfanov} et~al.}{2003}]{Gilf2003}
{Gilfanov} M.,  {Revnivtsev} M.,   {Molkov} S.,  2003, \mn@doi [\aap]
  {10.1051/0004-6361:20031141}, \href
  {https://ui.adsabs.harvard.edu/abs/2003A&A...410..217G} {410, 217}

\bibitem[\protect\citeauthoryear{{Glatzel}}{{Glatzel}}{1988}]{Glatzel1988}
{Glatzel} W.,  1988, \mn@doi [\mnras] {10.1093/mnras/231.3.795}, \href
  {https://ui.adsabs.harvard.edu/abs/1988MNRAS.231..795G} {231, 795}

\bibitem[\protect\citeauthoryear{{Goldreich} \& {Tremaine}}{{Goldreich} \&
  {Tremaine}}{1980}]{GT80}
{Goldreich} P.,  {Tremaine} S.,  1980, \mn@doi [\apj] {10.1086/158356}, \href
  {http://adsabs.harvard.edu/abs/1980ApJ...241..425G} {241, 425}

\bibitem[\protect\citeauthoryear{{Heinemann} \& {Papaloizou}}{{Heinemann} \&
  {Papaloizou}}{2009}]{Heinemann2009}
{Heinemann} T.,  {Papaloizou} J.~C.~B.,  2009, \mn@doi [\mnras]
  {10.1111/j.1365-2966.2009.14799.x}, \href
  {https://ui.adsabs.harvard.edu/abs/2009MNRAS.397...52H} {397, 52}

\bibitem[\protect\citeauthoryear{{Hertfelder} \& {Kley}}{{Hertfelder} \&
  {Kley}}{2015a}]{HER15}
{Hertfelder} M.,  {Kley} W.,  2015a, \mn@doi [\aap]
  {10.1051/0004-6361/201526005}, \href
  {http://adsabs.harvard.edu/abs/2015A%26A...579A..54H} {579, A54}

\bibitem[\protect\citeauthoryear{{Hertfelder} \& {Kley}}{{Hertfelder} \&
  {Kley}}{2015b}]{Hert2015}
{Hertfelder} M.,  {Kley} W.,  2015b, \mn@doi [\aap]
  {10.1051/0004-6361/201526005}, \href
  {https://ui.adsabs.harvard.edu/abs/2015A&A...579A..54H} {579, A54}

\bibitem[\protect\citeauthoryear{{Hertfelder} \& {Kley}}{{Hertfelder} \&
  {Kley}}{2017}]{Hert2017}
{Hertfelder} M.,  {Kley} W.,  2017, \mn@doi [\aap]
  {10.1051/0004-6361/201730847}, \href
  {https://ui.adsabs.harvard.edu/abs/2017A&A...605A..24H} {605, A24}

\bibitem[\protect\citeauthoryear{{Hertfelder}, {Kley}, {Suleimanov}  \&
  {Werner}}{{Hertfelder} et~al.}{2013}]{HER13}
{Hertfelder} M.,  {Kley} W.,  {Suleimanov} V.,   {Werner} K.,  2013, \mn@doi
  [\aap] {10.1051/0004-6361/201322542}, \href
  {http://adsabs.harvard.edu/abs/2013A%26A...560A..56H} {560, A56}

\bibitem[\protect\citeauthoryear{{Hirose}}{{Hirose}}{2015}]{HIR15}
{Hirose} S.,  2015, \mn@doi [\mnras] {10.1093/mnras/stv203}, \href
  {http://adsabs.harvard.edu/abs/2015MNRAS.448.3105H} {448, 3105}

\bibitem[\protect\citeauthoryear{{Hirose}, {Blaes}, {Krolik}, {Coleman}  \&
  {Sano}}{{Hirose} et~al.}{2014}]{HIR14}
{Hirose} S.,  {Blaes} O.,  {Krolik} J.~H.,  {Coleman} M.~S.~B.,   {Sano} T.,
  2014, \mn@doi [\apj] {10.1088/0004-637X/787/1/1}, \href
  {http://adsabs.harvard.edu/abs/2014ApJ...787....1H} {787, 1}

\bibitem[\protect\citeauthoryear{{Inogamov} \& {Sunyaev}}{{Inogamov} \&
  {Sunyaev}}{1999}]{INO99}
{Inogamov} N.~A.,  {Sunyaev} R.~A.,  1999, Astronomy Letters, \href
  {http://adsabs.harvard.edu/abs/1999AstL...25..269I} {25, 269}

\bibitem[\protect\citeauthoryear{{Inogamov} \& {Sunyaev}}{{Inogamov} \&
  {Sunyaev}}{2010}]{INO10}
{Inogamov} N.~A.,  {Sunyaev} R.~A.,  2010, \mn@doi [Astronomy Letters]
  {10.1134/S1063773710120029}, \href
  {http://adsabs.harvard.edu/abs/2010AstL...36..848I} {36, 848}

\bibitem[\protect\citeauthoryear{{Johnson} \& {Gammie}}{{Johnson} \&
  {Gammie}}{2005}]{Johnson2005}
{Johnson} B.~M.,  {Gammie} C.~F.,  2005, \mn@doi [\apj] {10.1086/497358}, \href
  {https://ui.adsabs.harvard.edu/abs/2005ApJ...635..149J} {635, 149}

\bibitem[\protect\citeauthoryear{{Kevlahan}}{{Kevlahan}}{1997}]{Kevlahan1997}
{Kevlahan} N.~K.~R.,  1997, \mn@doi [Journal of Fluid Mechanics]
  {10.1017/S0022112097005752}, \href
  {https://ui.adsabs.harvard.edu/abs/1997JFM...341..371K} {341, 371}

\bibitem[\protect\citeauthoryear{{Kippenhahn} \& {Thomas}}{{Kippenhahn} \&
  {Thomas}}{1978}]{Kip1978}
{Kippenhahn} R.,  {Thomas} H.~C.,  1978, \aap, \href
  {https://ui.adsabs.harvard.edu/abs/1978A&A....63..265K} {63, 265}

\bibitem[\protect\citeauthoryear{{Kley} \& {Lin}}{{Kley} \&
  {Lin}}{1996}]{Kley1996}
{Kley} W.,  {Lin} D.~N.~C.,  1996, \mn@doi [\apj] {10.1086/177115}, \href
  {https://ui.adsabs.harvard.edu/abs/1996ApJ...461..933K} {461, 933}

\bibitem[\protect\citeauthoryear{{Koenigl}}{{Koenigl}}{1991}]{KON91}
{Koenigl} A.,  1991, \mn@doi [\apjl] {10.1086/185972}, \href
  {http://adsabs.harvard.edu/abs/1991ApJ...370L..39K} {370, L39}

\bibitem[\protect\citeauthoryear{{Koller}, {Li}  \& {Lin}}{{Koller}
  et~al.}{2003}]{Koller2003}
{Koller} J.,  {Li} H.,   {Lin} D. N.~C.,  2003, \mn@doi [\apjl]
  {10.1086/379032}, \href
  {https://ui.adsabs.harvard.edu/abs/2003ApJ...596L..91K} {596, L91}

\bibitem[\protect\citeauthoryear{{Lesur} \& {Papaloizou}}{{Lesur} \&
  {Papaloizou}}{2009}]{Lesur2009}
{Lesur} G.,  {Papaloizou} J.~C.~B.,  2009, \mn@doi [\aap]
  {10.1051/0004-6361/200811577}, \href
  {https://ui.adsabs.harvard.edu/abs/2009A&A...498....1L} {498, 1}

\bibitem[\protect\citeauthoryear{{Li}, {Finn}, {Lovelace}  \& {Colgate}}{{Li}
  et~al.}{2000}]{Li2000}
{Li} H.,  {Finn} J.~M.,  {Lovelace} R.~V.~E.,   {Colgate} S.~A.,  2000, \mn@doi
  [\apj] {10.1086/308693}, \href
  {https://ui.adsabs.harvard.edu/abs/2000ApJ...533.1023L} {533, 1023}

\bibitem[\protect\citeauthoryear{{Li}, {Colgate}, {Wendroff}  \& {Liska}}{{Li}
  et~al.}{2001}]{Li2001}
{Li} H.,  {Colgate} S.~A.,  {Wendroff} B.,   {Liska} R.,  2001, \mn@doi [\apj]
  {10.1086/320241}, \href
  {https://ui.adsabs.harvard.edu/abs/2001ApJ...551..874L} {551, 874}

\bibitem[\protect\citeauthoryear{{Li}, {Li}, {Koller}, {Wendroff}, {Liska},
  {Orban}, {Liang}  \& {Lin}}{{Li} et~al.}{2005}]{Li2005}
{Li} H.,  {Li} S.,  {Koller} J.,  {Wendroff} B.~B.,  {Liska} R.,  {Orban}
  C.~M.,  {Liang} E. P.~T.,   {Lin} D. N.~C.,  2005, \mn@doi [\apj]
  {10.1086/429367}, \href
  {https://ui.adsabs.harvard.edu/abs/2005ApJ...624.1003L} {624, 1003}

\bibitem[\protect\citeauthoryear{{Lin}}{{Lin}}{2012}]{Lin2012}
{Lin} M.-K.,  2012, \mn@doi [\mnras] {10.1111/j.1365-2966.2012.21955.x}, \href
  {https://ui.adsabs.harvard.edu/abs/2012MNRAS.426.3211L} {426, 3211}

\bibitem[\protect\citeauthoryear{{Lin}}{{Lin}}{2015}]{Lin2015}
{Lin} M.-K.,  2015, \mn@doi [\mnras] {10.1093/mnras/stv254}, \href
  {http://adsabs.harvard.edu/abs/2015MNRAS.448.3806L} {448, 3806}

\bibitem[\protect\citeauthoryear{{Lin} \& {Papaloizou}}{{Lin} \&
  {Papaloizou}}{2010}]{LinPap2010}
{Lin} M.-K.,  {Papaloizou} J. C.~B.,  2010, \mn@doi [\mnras]
  {10.1111/j.1365-2966.2010.16560.x}, \href
  {https://ui.adsabs.harvard.edu/abs/2010MNRAS.405.1473L} {405, 1473}

\bibitem[\protect\citeauthoryear{{Lin} \& {Pierens}}{{Lin} \&
  {Pierens}}{2018}]{Lin2018}
{Lin} M.-K.,  {Pierens} A.,  2018, \mn@doi [\mnras] {10.1093/mnras/sty947},
  \href {https://ui.adsabs.harvard.edu/abs/2018MNRAS.478..575L} {478, 575}

\bibitem[\protect\citeauthoryear{{Lithwick}}{{Lithwick}}{2009}]{Lithwick2009}
{Lithwick} Y.,  2009, \mn@doi [\apj] {10.1088/0004-637X/693/1/85}, \href
  {https://ui.adsabs.harvard.edu/abs/2009ApJ...693...85L} {693, 85}

\bibitem[\protect\citeauthoryear{{Lovelace}, {Li}, {Colgate}  \&
  {Nelson}}{{Lovelace} et~al.}{1999}]{Lovelace1999}
{Lovelace} R.~V.~E.,  {Li} H.,  {Colgate} S.~A.,   {Nelson} A.~F.,  1999,
  \mn@doi [\apj] {10.1086/306900}, \href
  {https://ui.adsabs.harvard.edu/abs/1999ApJ...513..805L} {513, 805}

\bibitem[\protect\citeauthoryear{{Meheut}, {Yu}  \& {Lai}}{{Meheut}
  et~al.}{2012a}]{Meheut2012b}
{Meheut} H.,  {Yu} C.,   {Lai} D.,  2012a, \mn@doi [\mnras]
  {10.1111/j.1365-2966.2012.20789.x}, \href
  {https://ui.adsabs.harvard.edu/abs/2012MNRAS.422.2399M} {422, 2399}

\bibitem[\protect\citeauthoryear{{Meheut}, {Keppens}, {Casse}  \&
  {Benz}}{{Meheut} et~al.}{2012b}]{Meheut2012a}
{Meheut} H.,  {Keppens} R.,  {Casse} F.,   {Benz} W.,  2012b, \mn@doi [\aap]
  {10.1051/0004-6361/201118500}, \href
  {https://ui.adsabs.harvard.edu/abs/2012A&A...542A...9M} {542, A9}

\bibitem[\protect\citeauthoryear{{Miranda} \& {Rafikov}}{{Miranda} \&
  {Rafikov}}{2019a}]{MR19}
{Miranda} R.,  {Rafikov} R.~R.,  2019a, \mn@doi [\apj]
  {10.3847/1538-4357/ab0f9e}, \href
  {https://ui.adsabs.harvard.edu/abs/2019ApJ...875...37M} {875, 37}

\bibitem[\protect\citeauthoryear{{Miranda} \& {Rafikov}}{{Miranda} \&
  {Rafikov}}{2019b}]{Miranda-ALMA}
{Miranda} R.,  {Rafikov} R.~R.,  2019b, \mn@doi [The Astrophysical Journal]
  {10.3847/2041-8213/ab22a7}, \href
  {https://ui.adsabs.harvard.edu/abs/2019ApJ...878L...9M} {878, L9}

\bibitem[\protect\citeauthoryear{{Miranda} \& {Rafikov}}{{Miranda} \&
  {Rafikov}}{2020}]{Miranda-Cooling}
{Miranda} R.,  {Rafikov} R.~R.,  2020, \mn@doi [\apj]
  {10.3847/1538-4357/ab791a}, \href
  {https://ui.adsabs.harvard.edu/abs/2020ApJ...892...65M} {892, 65}

\bibitem[\protect\citeauthoryear{{Narayan} \& {Popham}}{{Narayan} \&
  {Popham}}{1993}]{NAR93}
{Narayan} R.,  {Popham} R.,  1993, \mn@doi [\nat] {10.1038/362820a0}, \href
  {http://adsabs.harvard.edu/abs/1993Natur.362..820N} {362, 820}

\bibitem[\protect\citeauthoryear{{Narayan}, {Goldreich}  \&
  {Goodman}}{{Narayan} et~al.}{1987}]{NGG1987}
{Narayan} R.,  {Goldreich} P.,   {Goodman} J.,  1987, \mn@doi [\mnras]
  {10.1093/mnras/228.1.1}, \href
  {https://ui.adsabs.harvard.edu/abs/1987MNRAS.228....1N} {228, 1}

\bibitem[\protect\citeauthoryear{{Ogilvie} \& {Lubow}}{{Ogilvie} \&
  {Lubow}}{2002}]{OL02}
{Ogilvie} G.~I.,  {Lubow} S.~H.,  2002, \mn@doi [\mnras]
  {10.1046/j.1365-8711.2002.05148.x}, \href
  {https://ui.adsabs.harvard.edu/abs/2002MNRAS.330..950O} {330, 950}

\bibitem[\protect\citeauthoryear{{Ono}, {Muto}, {Takeuchi}  \& {Nomura}}{{Ono}
  et~al.}{2016}]{Ono2016}
{Ono} T.,  {Muto} T.,  {Takeuchi} T.,   {Nomura} H.,  2016, \mn@doi [\apj]
  {10.3847/0004-637X/823/2/84}, \href
  {https://ui.adsabs.harvard.edu/abs/2016ApJ...823...84O} {823, 84}

\bibitem[\protect\citeauthoryear{{Ono}, {Muto}, {Tomida}  \& {Zhu}}{{Ono}
  et~al.}{2018}]{Ono2018}
{Ono} T.,  {Muto} T.,  {Tomida} K.,   {Zhu} Z.,  2018, \mn@doi [\apj]
  {10.3847/1538-4357/aad54d}, \href
  {https://ui.adsabs.harvard.edu/abs/2018ApJ...864...70O} {864, 70}

\bibitem[\protect\citeauthoryear{{Paardekooper}, {Lesur}  \&
  {Papaloizou}}{{Paardekooper} et~al.}{2010}]{Paar2010}
{Paardekooper} S.-J.,  {Lesur} G.,   {Papaloizou} J. C.~B.,  2010, \mn@doi
  [\apj] {10.1088/0004-637X/725/1/146}, \href
  {https://ui.adsabs.harvard.edu/abs/2010ApJ...725..146P} {725, 146}

\bibitem[\protect\citeauthoryear{{Papaloizou} \& {Lin}}{{Papaloizou} \&
  {Lin}}{1989}]{Pap1989}
{Papaloizou} J.~C.~B.,  {Lin} D.~N.~C.,  1989, \mn@doi [\apj] {10.1086/167832},
  \href {https://ui.adsabs.harvard.edu/abs/1989ApJ...344..645P} {344, 645}

\bibitem[\protect\citeauthoryear{{Papaloizou} \& {Pringle}}{{Papaloizou} \&
  {Pringle}}{1984}]{PP1984}
{Papaloizou} J.~C.~B.,  {Pringle} J.~E.,  1984, \mn@doi [\mnras]
  {10.1093/mnras/208.4.721}, \href
  {https://ui.adsabs.harvard.edu/abs/1984MNRAS.208..721P} {208, 721}

\bibitem[\protect\citeauthoryear{{Pessah} \& {Chan}}{{Pessah} \&
  {Chan}}{2012}]{Pes2012}
{Pessah} M.~E.,  {Chan} C.-k.,  2012, \mn@doi [\apj]
  {10.1088/0004-637X/751/1/48}, \href
  {https://ui.adsabs.harvard.edu/abs/2012ApJ...751...48P} {751, 48}

\bibitem[\protect\citeauthoryear{{Philippov}, {Rafikov}  \&
  {Stone}}{{Philippov} et~al.}{2016}]{PHI16}
{Philippov} A.~A.,  {Rafikov} R.~R.,   {Stone} J.~M.,  2016, \mn@doi [\apj]
  {10.3847/0004-637X/817/1/62}, \href
  {http://adsabs.harvard.edu/abs/2016ApJ...817...62P} {817, 62}

\bibitem[\protect\citeauthoryear{{Popham}, {Narayan}, {Hartmann}  \&
  {Kenyon}}{{Popham} et~al.}{1993}]{POP93}
{Popham} R.,  {Narayan} R.,  {Hartmann} L.,   {Kenyon} S.,  1993, \mn@doi
  [\apjl] {10.1086/187049}, \href
  {http://adsabs.harvard.edu/abs/1993ApJ...415L.127P} {415, L127}

\bibitem[\protect\citeauthoryear{{Rafikov}}{{Rafikov}}{2002}]{R02}
{Rafikov} R.~R.,  2002, \mn@doi [\apj] {10.1086/339399}, \href
  {http://adsabs.harvard.edu/abs/2002ApJ...569..997R} {569, 997}

\bibitem[\protect\citeauthoryear{{Revnivtsev} \& {Gilfanov}}{{Revnivtsev} \&
  {Gilfanov}}{2006}]{Rev2006}
{Revnivtsev} M.~G.,  {Gilfanov} M.~R.,  2006, \mn@doi [\aap]
  {10.1051/0004-6361:20053964}, \href
  {https://ui.adsabs.harvard.edu/abs/2006A&A...453..253R} {453, 253}

\bibitem[\protect\citeauthoryear{{Roelofs}, {Groot}, {Benedict}, {McArthur},
  {Steeghs}, {Morales-Rueda}, {Marsh}  \& {Nelemans}}{{Roelofs}
  et~al.}{2007}]{2007ApJ...666.1174R}
{Roelofs} G.~H.~A.,  {Groot} P.~J.,  {Benedict} G.~F.,  {McArthur} B.~E.,
  {Steeghs} D.,  {Morales-Rueda} L.,  {Marsh} T.~R.,   {Nelemans} G.,  2007,
  \mn@doi [\apj] {10.1086/520491}, \href
  {https://ui.adsabs.harvard.edu/abs/2007ApJ...666.1174R} {666, 1174}

\bibitem[\protect\citeauthoryear{{Shakura} \& {Sunyaev}}{{Shakura} \&
  {Sunyaev}}{1973}]{SS73}
{Shakura} N.~I.,  {Sunyaev} R.~A.,  1973, \aap, \href
  {http://adsabs.harvard.edu/abs/1973A%26A....24..337S} {24, 337}

\bibitem[\protect\citeauthoryear{{Steinacker} \& {Papaloizou}}{{Steinacker} \&
  {Papaloizou}}{2002}]{Stein2002}
{Steinacker} A.,  {Papaloizou} J. C.~B.,  2002, \mn@doi [\apj]
  {10.1086/339892}, \href
  {https://ui.adsabs.harvard.edu/abs/2002ApJ...571..413S} {571, 413}

\bibitem[\protect\citeauthoryear{{Stone}, {Tomida}, {White}  \&
  {Felker}}{{Stone} et~al.}{2020}]{athena}
{Stone} J.~M.,  {Tomida} K.,  {White} C.~J.,   {Felker} K.~G.,  2020, \mn@doi
  [\apjs] {10.3847/1538-4365/ab929b}, \href
  {https://ui.adsabs.harvard.edu/abs/2020ApJS..249....4S} {249, 4}

\bibitem[\protect\citeauthoryear{Towns et~al.,}{Towns et~al.}{2014}]{XSEDE}
Towns J.,  et~al., 2014, \mn@doi [Computing in Science & Engineering]
  {10.1109/MCSE.2014.80}, 16, 62

\bibitem[\protect\citeauthoryear{{Velikhov}}{{Velikhov}}{1959}]{VEL59}
{Velikhov} E.~P.,  1959, Journal of Experimental and Theoretical Physics, 36,
  995

\bibitem[\protect\citeauthoryear{{Warner}}{{Warner}}{2003}]{WAR03}
{Warner} B.,  2003, {Cataclysmic Variable Stars},
  \mn@doi{10.1017/CB09780511586491.
}

\bibitem[\protect\citeauthoryear{{de Val-Borro}, {Artymowicz}, {D'Angelo}  \&
  {Peplinski}}{{de Val-Borro} et~al.}{2007}]{deVal2007}
{de Val-Borro} M.,  {Artymowicz} P.,  {D'Angelo} G.,   {Peplinski} A.,  2007,
  \mn@doi [\aap] {10.1051/0004-6361:20077169}, \href
  {https://ui.adsabs.harvard.edu/abs/2007A&A...471.1043D} {471, 1043}

\makeatother
\end{thebibliography}



\appendix


\section{Dispersion relation for the upper modes}
\label{sect:upper-DR}


\citet{BRS13a} derived a dispersion relation for modes with arbitrary degree of winding, given by their Eqn.~(A16). It assumes globally isothermal equation of state and constant $\Sigma$ in the disk. The degree of winding is quantified by the parameter $n=r k_r$; for tightly wound waves $n\gg 1$ and one recovers the usual WKB relation (\ref{eq:disp_rel}). However, for the upper modes $k_r\to 0$ as $r\to R_\star$, so that $n=0$. In this limit the dispersion relation (A16) of \citet{BRS13a} becomes 
\ba  
\left[\Omega(R_\star)-\Omega_{\rm p}\right]^2=\frac{c_s^2}{R_\star^2}+\frac{\kappa^2(R_\star)}{m^2}-\frac{2}{m^2} \frac{c_s^2}{R_\star^2}\frac{\Omega(R_\star)}{\Omega(R_\star)-\Omega_{\rm p}},
\label{eq:refinedDR}
\ea  
where we also set $r\to R_\star$; in the limit $m\gg 1$ we recover equation (\ref{eq:DR-uppr-approx}). Note that the relation (\ref{eq:refinedDR}) is real, whereas for general $n\neq 0$ the Eqn. (A16) of \citet{BRS13a} is not. 

Equation (\ref{eq:refinedDR}) is a cubic in $\Omega_{\rm p}$, which can be solved for $\Omega_{\rm p}(m)$. However, a simpler procedure is to solve for $m(\Omega_{\rm p})$:
\ba
    m^2=\frac{\kappa^2\left(r_\mathrm{f}\right)r_\mathrm{f}^{2}\left[\Omega\left(r_\mathrm{f}\right)-\Omega_{\rm p}\right]-2 c_s^2 \Omega\left(r_\mathrm{f}\right)}{ r_\mathrm{f}^{2}\left[\Omega\left(r_\mathrm{f}\right)-\Omega_{\rm p}\right]^3-c_s^2\left[\Omega\left(r_\mathrm{f}\right)-\Omega_{\rm p}\right]},
\label{eqn:upper_disp1}
\ea
where in simulation units $c_{\rm s} =\M^{-1}$, and $r_\mathrm{f}$ is a fiducial radius chosen (instead of $R_\star$) to correspond to the maximum of $\Omega$ in the time- (for $t/2\pi>100$) and azimuthally-averaged simulation data. Typically $r_\mathrm{f}$ is very close to $R_\star$.

Note that at late times $\Omega(r)$ develops a plateau near $r=R_\star\approx r_\mathrm{f}$ because $\Sigma$ drops in this region, see Eqn. (\ref{eq:Omega}). 
As a result,  $\Omega(r_\mathrm{f})<\Omega_K(r_\mathrm{f})$ and  $\kappa(r_\mathrm{f})\neq \Omega_K(r_\mathrm{f})$. 
If $\Omega$ is truly constant over some radial interval near the BL then $\kappa(r_\mathrm{f})\approx 2\Omega(r_\mathrm{f})$. 
While this approximation is very accurate at $r_\mathrm{f}$ (where $\Omega$ takes its peak value) we still compute $\kappa$ directly from the time averaged simulation data to draw the upper mode dispersion relation in Figure \ref{fig:multi_dispersion} using the Eqn. (\ref{eqn:upper_disp1}).


\section{Details of the numerical analysis tools}
\label{sect:num_an}



\subsection{Runtime Fourier Analysis}
\label{sect:FFT}


To save on disk space we modified the FFT module of \athena (designed to solve Poisson's equation for self-gravity, and to inject turbulence) to perform FFT on the fluid variables in the azimuthal ($\varphi$) direction only and save the $m=0$ through 31 modes as a function of $r$ to file (saving us a factor of $32/n_\varphi$ in disk space). 
We save two separate sets of these FFT files at a cadence of $\Delta t=10^{-1}/2\pi$, resulting in 20 outputs per (inner) orbit. These two sets of FFT files are identical but temporally offset from each-other by $\delta t=10^{-2}/2\pi$. 
This enables us to perform finite differences to estimate the time derivative of the (complex) phase ($\varphi$) to order $\delta t\Delta t$. This offset strategy also saves disk space by enabling us to resolve $m\dot{\varphi}<1/\delta t$ without needing 100 FFT files per orbit. To further expedite post-processing we binned these results by orbit (using the mean data value within that bin). 
The standard deviation within each bin is used as an error estimate for that data. While the majority of our FFT diagnostics in this paper use the Fourier transform of $v_r\sqrt{\Sigma}$, the FFTs of several other variables are also computed.


\subsection{Automated Mode Detection}
\label{sec:mode_detect}


We make use of these FFT data sets to automatically detect stable modes in our runs. The first step in this process is to bin the FFT data into 10 orbit chunks (10 data points per chunk), making the data a function of $t$, $r$, $m$.
The first 30 orbits of data are discarded as they are sensitive to the initial conditions. We then focus on two radial locations, $r_1=(r_{\rm min}/2 + 1)/2$ and $r_2=1.2$ 
and preform an error weighted radial mean over $\pm 5$ radial grid cells about these radii of interest. 
Within each 10-orbit chunk we
preform linear regression of $\Omega_{\rm p}(t)$ (for each $m,\,r_i$). 
We discard data (by 10 orbit chunk) with mean values $\Omega_{p\rm }>1$ or $\Omega_{\rm p}<0$, and further filter out data with $|\partial \Omega_{\rm p} / \partial t| > (2\times 10^{-4}, 4\times 10^{-4})$, or standard deviation from the linear fit $> (1\times 10^{-4}, 2\times 10^{-4})$ for $r=(r_1,r_2)$ respectively.
Remaining data where three adjacent 10-orbit chunks have passed the filtering process are considered stable modes. If the same modes are detected at all considered radii, with overlapping times, they are classified as global modes.


\subsection{Details of the time-averaging procedures in mode figures}
\label{sect:averaging}


Here we summarize the technical details of various averaging and smoothing procedures that were used in analyzing the data and producing various plots.
All of the colored lines in Figures \ref{fig:M09.FR.r.a-diag}, \ref{fig:M06.HR.r.a-diag}, \ref{fig:M12.FR.r.a-diag}, and \ref{fig:M15.FR.r.a-diag} correspond to a specific mode, which is shown in the same color in all four panels of a given figure. 
Five of these lines correspond to the five modes with the highest time (after the first 30 orbits) and space integrated amplitude. 
Where applicable, an additional sixth mode that is classified as a global mode at some point during the simulation is also drawn.
The data shown in these four figures are variance-weighted means (using the variance of a given quantity within a one-orbit bin; see Section \ref{sect:FFT}) of the data in the radial direction, $\pm5$ and $\pm30$ grid points for $r<1$ and $r>1$ respectively.
Finally, the data is convolved with a symmetric triangular window in the time domain of width 20 orbits. 
This window function is chosen for its simplicity and to suppress leakage of power to high frequencies in comparison to a rectangular window. 
We also ran a few tests to examine the impact of various window functions as saw no appreciable difference.


\section{Morphological characterization of wave structures for different $\M$}
\label{sect:otherMs}


Here we provide the description of the wave patterns emerging in our simulations for different values of $\M$, similar to \S\ref{sect:fiducialM=9}.


\subsection{Other $\M=9$ and similar runs}
\label{sect:other-M9}


In addition to the run M09.FR.r.a described in \S \ref{sect:fiducialM=9}, we have also carried out 14 more $\M=9$ runs to test the sensitivity of outcomes to our adopted resolution and initial conditions. With very few exceptions, all these runs show the behavior similar to M09.FR.r.a: a long period of the lower $m\approx 19-21$ mode dominance, almost always in conjunction with the $m=2$, $k_r=0$ pattern in the disk. In the last $200-300$ orbits of these runs we quite robustly find the large scale $m=4-6$ spiral arms in the disk (\S\ref{sect:vortex}) co-existing with the high $m\sim 20-28$ upper mode inside the star. Also, in more than $50\%$ of the runs, after $t/2\pi=300-400$, we see an $m=4-7$ disk mode confined to the vicinity of the stellar surface (\S\ref{sect:low-m_disc}) and superposed on the other modes. 

\begin{figure}
\ifLOWRES
	\includegraphics[width=.99\linewidth]{figs/lowres/lowres_M06.HR.r.a_maps_stripes_2.png}
\else
    \includegraphics[width=.99\linewidth]{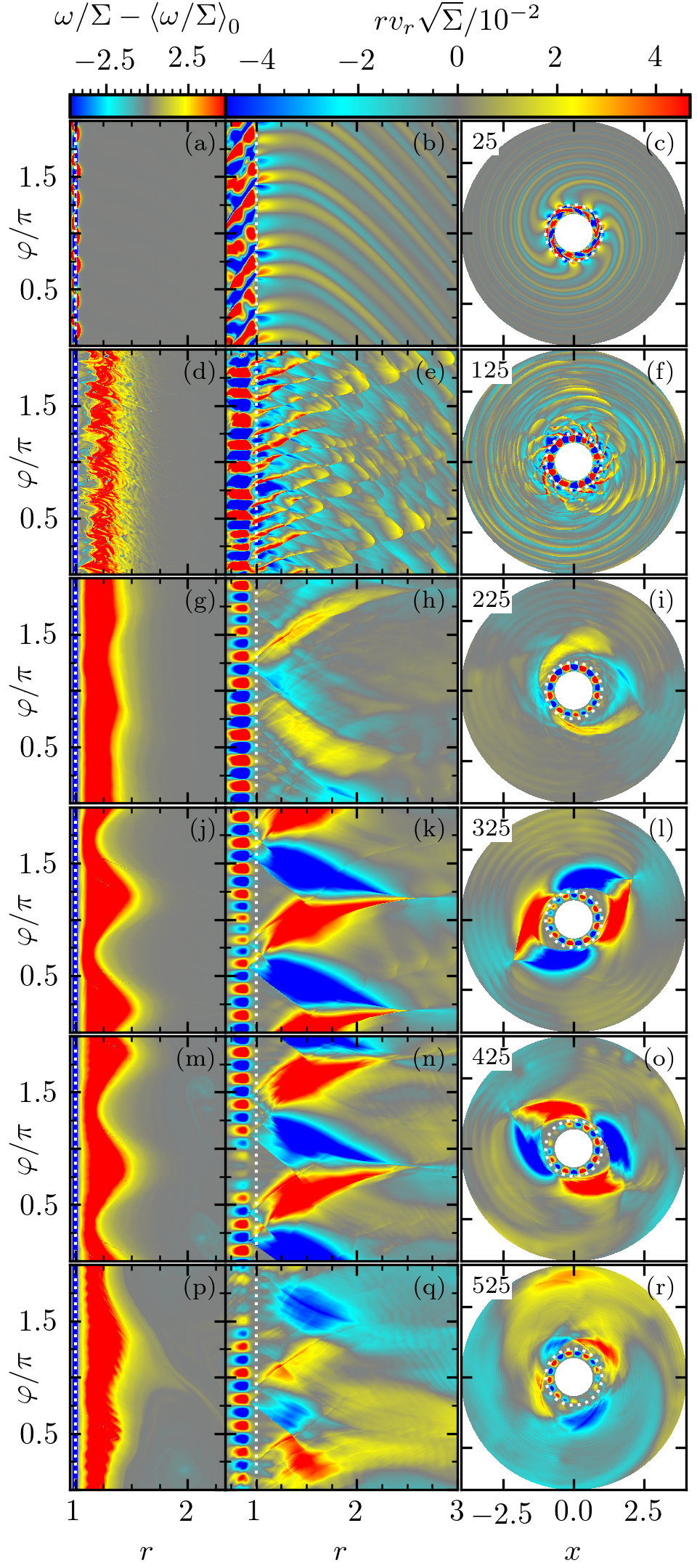}
\fi
    \caption{Same as Figure \ref{fig:M09.FR.r.a} but for the $\M=6$ run M06.HR.r.a (note a different radial range in the left and central columns chosen for better illustration of important features). Time corresponding to each snapshot is indicated at the top. See \S \ref{sect:M=6} for details.
    }
    \label{fig:M06.HR.r.a}
\end{figure}

\begin{figure}
	\includegraphics[width=\linewidth]{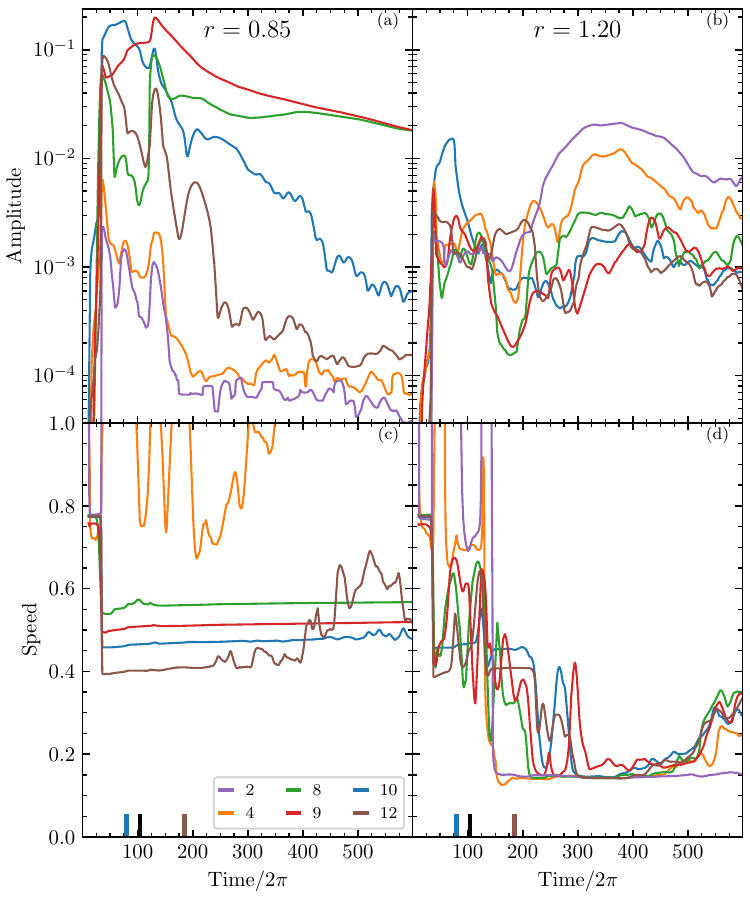}
    \caption{Same as Fig.~\ref{fig:M09.FR.r.a-diag}, but for the $\M=6$ simulation M06.HR.r.a. 
    }
    \label{fig:M06.HR.r.a-diag}
\end{figure}

We also ran one of our $\M=9$ simulations for much longer than the nominal duration of 600 orbit, up to $t/2\pi=3780$, to explore the long-term evolution of the disk+star system. Up to $t/2\pi=600$ the evolution followed the usual pattern of other $\M=9$ runs, and then the system settled into a state (lasting until $t/2\pi\sim 3000$) with the high-$m$ upper mode dominating inside the star, whereas an $m=3-4$ mode trapped near the stellar surface (to be discussed in \S\ref{sect:low-m_disc}) dominated in the disk. Then another re-arrangement happens, with a sequence of mergers of several global spirals into a single narrow one-armed spiral, followed by the development of a strong, stable $m=13$ lower mode. This is how this long simulation ends, giving us a hint of the complicated long-term evolution mediated by the instabilities in the BL. 

We have also carried out $\M=10$ and $11$ runs with the same initial conditions as the $\M=9$ run described in \S \ref{sect:fiducialM=9}. They show an evolutionary sequence  very similar to that in $\M=9$ runs. The only notable feature worth mentioning is the appearance of a small number of {\it narrow} spiral arms in the beginning of the $\M=11$ run, illustrated in Figure \ref{fig:vortex-types}; their origin is discussed in \S \ref{sect:vortex}.  


\subsection{$\M=6$ and other low-$\M$ runs.}
\label{sect:M=6}


Some features of the behavior of the disk-star system for $\M=6$ have been discussed in previous studies \citep{BRS12,BRS13a}, although those simulations were typically not advanced for as long as our current runs. Here we describe the details of a M06.HR.r.a run performed at the resolution of $2048\times 2048$. 

In the beginning, at $t/2\pi=25$, this run features a strong upper $m=7$ mode, see Figure \ref{fig:M06.HR.r.a}a. This mode is rather short lived and goes away already by $t/2\pi=50$, yielding to the lower mode with $m=10-11$ and $\Omega_{\rm p}\approx 0.47$. In the disk it can be traced until $t/2\pi\lesssim 200$ in the form of criss-crossing leading and trailing spirals confined to a resonant cavity, $r\lesssim 1.7-2$. Multiple shocks associated with this mode rapidly drive a substantial re-arrangement of the disk $\Sigma$ and $\Omega$ profile near the star, similar to  Figure \ref{fig:M09.FR.r.a_evo_prof}. Inside the star this lower mode dominates until the end of the run, with $m$ slowly evolving from 11 to 9, see Figures \ref{fig:M06.HR.r.a}b \& \ref{fig:M06.HR.r.a-diag}.  This evolution is in agreement with  \citet{BRS12,BRS13a}.

Around $t/2\pi=225$ (Figure \ref{fig:M06.HR.r.a}c) a new feature emerges: in the disk an $m=2$ mode replaces the lower $m=9-10$ mode (which still persists in the star). The low $\Omega_{\rm p}\approx 0.15$ of this mode (Figure \ref{fig:M06.HR.r.a-diag}d) places its corotation radius at $r_\mathrm{c}\approx 3.5$, while its ILR is at $r_\mathrm{ILR}\approx 2.3$, see equation (\ref{eq:r_ILR}). Most of the mode power is concentrated at $r\lesssim 2.5$, which is compatible with it being trapped interior to the $r_\mathrm{ILR}$. Despite its radial confinement close to the star, this mode is not a lower mode: its location in the dispersion relation (an orange cross in the lower left corner of Figure \ref{fig:multi_dispersion}b) is far from the lower mode branch. In fact, it falls near the upper mode branch of the dispersion relation, but this mode's morphology is also incompatible with the classical upper mode behavior, e.g. its $k_r(r)$ is non-zero as $r\to 1$. This mode is discussed in more detail in \S\ref{sect:low-m_disc}, where we show that it has the same nature as the $m=6$ mode confined to the inner disk in the $\M=9$ run described in \S\ref{sect:fiducial}. Similar persistent low $m=2-3$ patterns in the disk have been previously reported in \citet{BRS12}. The $m=2$ mode perists almost until the end of the simulation, when it becomes more chaotic and a strong $m=1$ perturbation (discussed in \S \ref{sect:vortex}) develops on top of it. 

Other runs carried out for similar values of $\M$ ranging from $5$ to $8$ paint a picture similar to the $\M=6$ case: lower modes dominating inside the star and the trapped low-$m$ modes (e.g. $m=2$ for $\M=5$ and $m=4$ for $\M=8$) dominating in the disk, typically with very low $\Omega_{\rm p}$ allowing them to extend far out. 

An interesting exception is the $\M=7$ case, in which the disk is dominated for the majority of the run by a small number, $2-5$, of global spiral arms (see Figure \ref{fig:vortex-types}d), similar to what is observed in the end of $\M=9$ run described in \S \ref{sect:fiducialM=9}. These modes may look like upper modes, but their $\Omega_{\rm p}\approx 0.75$ is too high for an upper mode (which can be deduced from Figure \ref{fig:multi_dispersion}c); their nature will be addressed in \S\ref{sect:vortex}. This run also features a superposed prominent $m=16$ lower mode (twice the $m$ of the lower mode dominating inside the star) locked in the resonant cavity $1<r\lesssim 1.8$, which makes the $\M=7$ run look even more distinct from $\M=6$ or $8$ simulations. These observations demonstrate that a particular mix of modes emerging in simulations is not a strictly monotonic function of $\M$.   
\FloatBarrier

\begin{figure}
\ifLOWRES
	\includegraphics[width=0.99\linewidth]{figs/lowres/{}lowres_M12.FR.r.a_maps_stripes_2.png}
\else
    \includegraphics[width=0.99\linewidth]{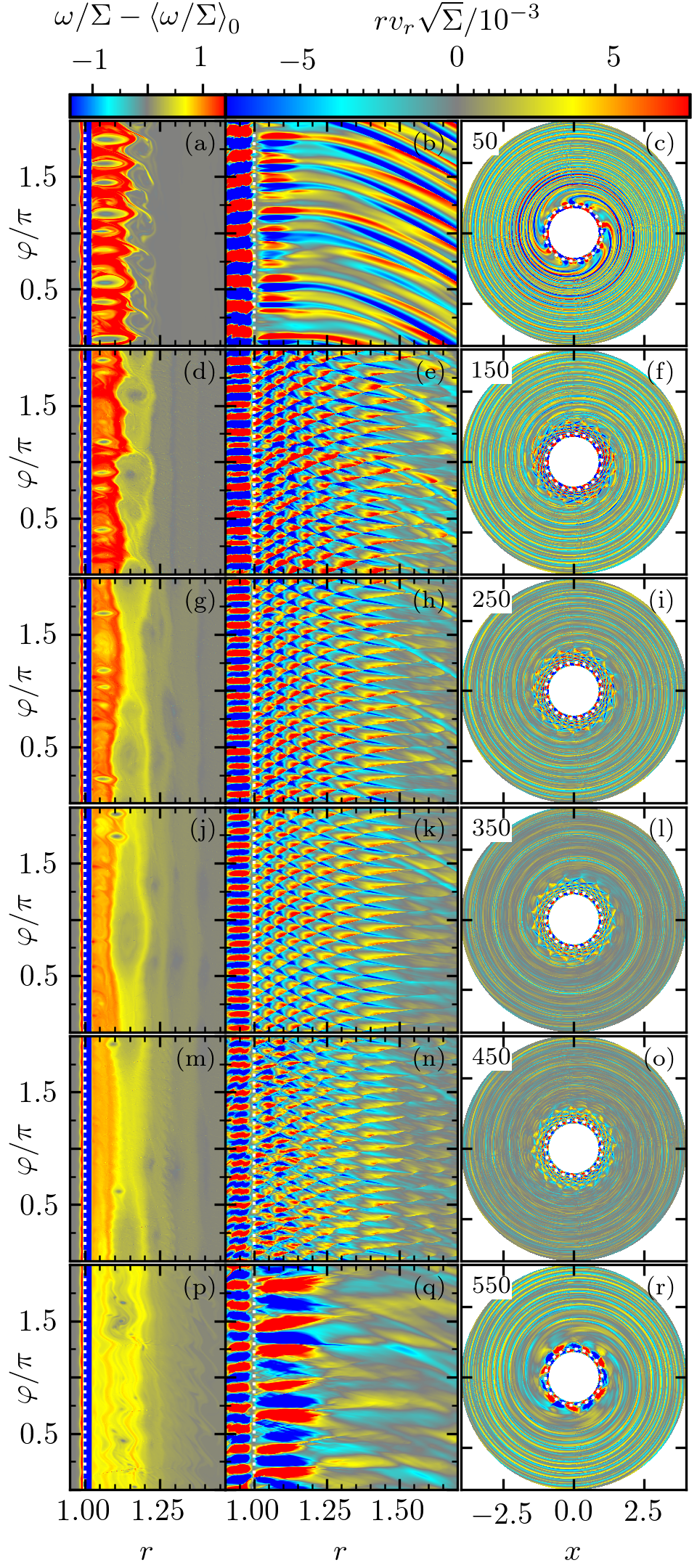}
\fi
    \caption{Same as Fig.~\ref{fig:M09.FR.r.a} but for the $\M=12$ simulation M12.FR.r.a. Note a very different radial extent ($r<1.7$) of the Cartesian maps chosen to illustrate the most important structures. See \S \ref{sect:M=12} for details.
    }
    \label{fig:M12.FR.r.a}
\end{figure}

\begin{figure}
	\includegraphics[width=\linewidth]{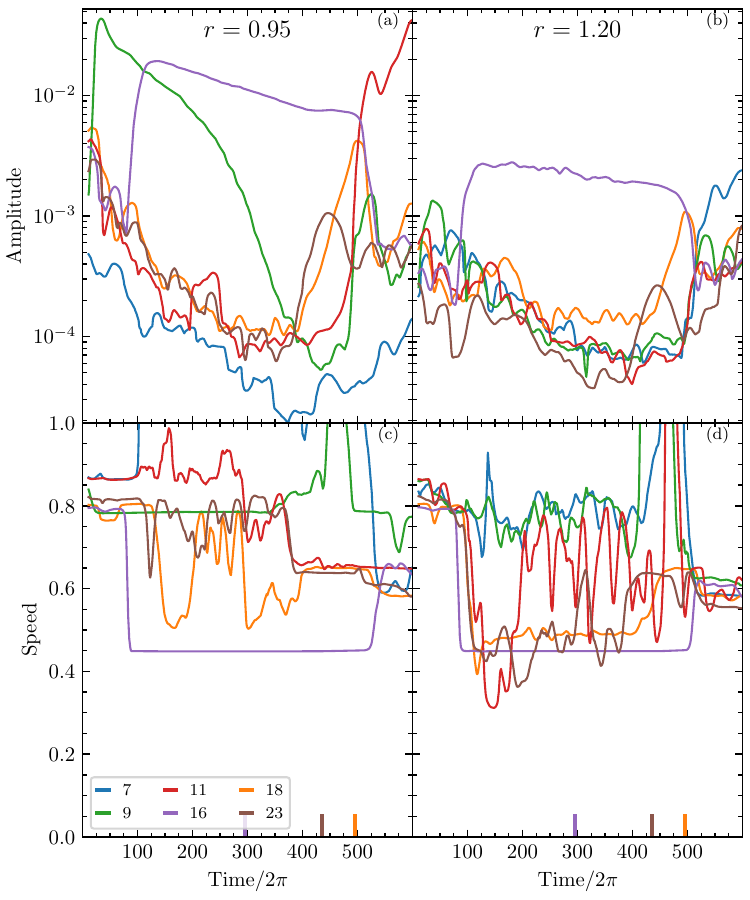}
    \caption{Same as Fig.~\ref{fig:M09.FR.r.a-diag}, but for the $\M=12$ simulation M12.FR.r.a.
    }
    \label{fig:M12.FR.r.a-diag}
\end{figure}


\subsection{$\M=12$ runs.}
\label{sect:M=12}


Runs at Mach number $\M=12$ (performed at the resolution of $4096\times 4096$) show surprisingly little variation in the outcomes, regardless of the initial conditions. Below we will briefly describe the evolution of the system in the M12.FR.r.a simulation.

In the beginning (at $t/2\pi=50$) a lower $m=9$ mode dominates inside the star, see Figure \ref{fig:M12.FR.r.a}. At the same time, in the disk we see multiple global spiral arms, which are atypical for lower modes but are very similar in appearance to the disk modes present in the beginning of the $\M=9$ run, see Figure \ref{fig:M09.FR.r.a}Ab,Ac and \S \ref{sect:vortex}. 

By $t/2\pi=125-150$ perturbation pattern changes and the dominance switches to the global lower $m=16$ mode. This mode has a low $\Omega_{\rm p}\approx 0.45$, which guarantees a substantial radial extent of the resonant cavity in which the mode is trapped: its corotation radius is $r_\mathrm{c}\approx 1.7$. This mode is very stable and persists roughly until $t/2\pi=450$, see Figure \ref{fig:M12.FR.r.a}. 

Beyond that point, by 550 orbits, the lower $m=16$ mode fades away, and the dominant mode inside the star becomes the $m=11$ lower mode with $\Omega_{\rm p}\approx 0.65$ and ILR at $r_\mathrm{ILR}\approx 1.25$. However, in the disk we do not find 11 criss-crossing waves confined between the star and $r_\mathrm{ILR}$, as one would expect for a global $m=11$ mode. Instead, we see an $m=7$ perturbation pattern confined to $1<r\lesssim 1.2$. This is quite reminiscent of the situation described in \S \ref{sect:M=6}, where we found the disk to feature an unusually low-$m$ mode ($m=2$ in that case) trapped near the stellar surface, simultaneously with a higher-$m$ lower mode inside the star. 

In addition to the $m=7$ mode trapped near the star, Figure \ref{fig:M12.FR.r.a}l,o,r also reveals presence of the large scale global spiral arms extending to the outer boundary of our domain. Although somewhat less coherent, these spirals are similar to the spiral arms found closer to the end of $\M=9$ run, see Figure \ref{fig:M09.FR.r.a}. They will be discussed further in \S \ref{sect:vortex}. 

Remarkably, all other $\M=12$ runs show very similar regular behavior, down to minor details: a dominant $m=9$ lower mode with high $\Omega_{\rm p}\approx 0.8$ in the beginning, changing around $t/2\pi=125-175$ to lower $m=16$ mode confined between $r=1$ and $r\approx 1.5-1.6$. Only the time at which $m=16$ lower mode fades away shows some variation, roughly between $t/2\pi=400$ and $550$. 

The regular and stable behavior exhibited by $\M=12$ runs is rather unique. The $\M=10,11$ runs exhibit the behavior which is much closer to $\M=9$ simulations. At the same time $\M=13,14$ runs are also very different from $\M=12$ ones, and resemble $\M=15$ run, see next. 

\FloatBarrier

\begin{figure}
\ifLOWRES
	\includegraphics[width=0.99\linewidth]{figs/lowres/{}lowres_M15.FR.r.a_maps_stripes_2.png}
\else
    \includegraphics[width=0.99\linewidth]{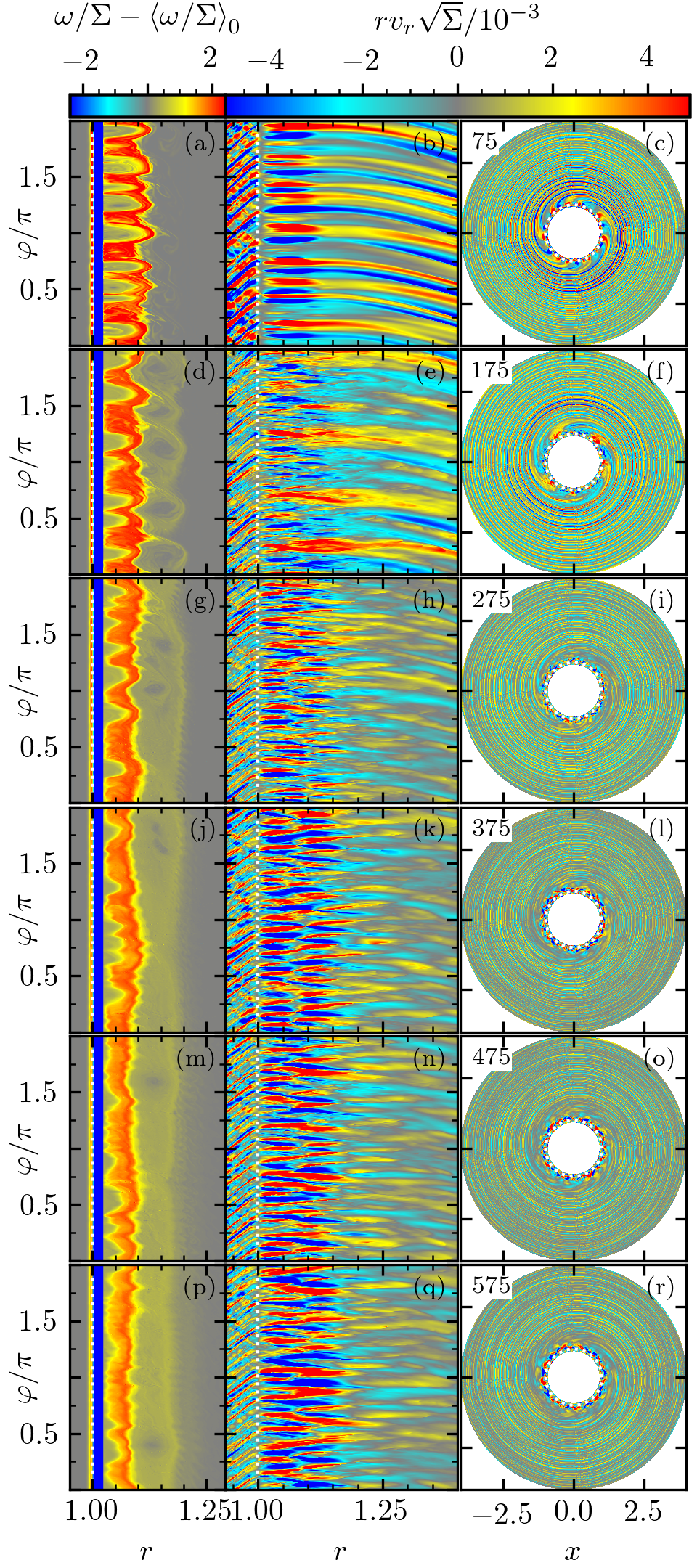}
\fi
    \caption{Same as Fig.~\ref{fig:M09.FR.r.a} but for the $\M=15$ simulation M15.FR.r.a. The radial range ($r<1.4$) is considerably smaller in Cartesian maps. See \S\ref{sect:M=15} for details.
    }
    \label{fig:M15.FR.r.a}
\end{figure}

\begin{figure}
	\includegraphics[width=\linewidth]{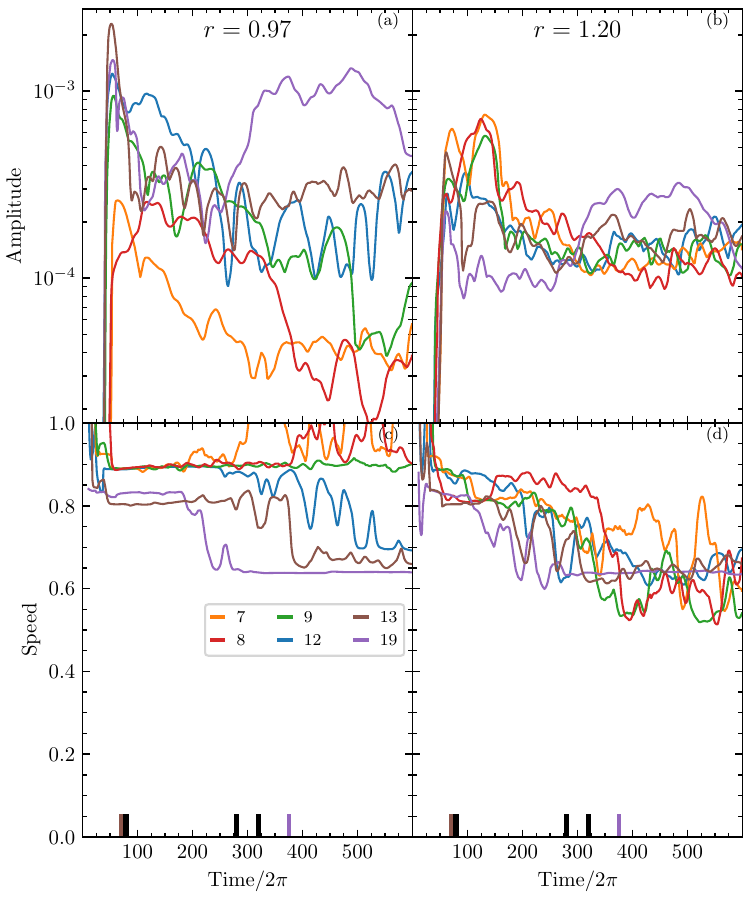}
    \caption{Same as Fig.~\ref{fig:M09.FR.r.a-diag}, but for the $\M=15$ simulation M15.FR.r.a.}
    \label{fig:M15.FR.r.a-diag}
\end{figure}


\subsection{$\M=13-15$ runs.}
\label{sect:M=15}


Simulations for $\M=13-15$ have resolution $8192\times 8192$ and initial conditions ('r.a') identical to those used in the $\M=6,9,12$ runs discussed earlier. We describe their outcomes using the $\M=15$ run M15.FR.r.a, see Figure \ref{fig:M15.FR.r.a}.

Pretty much all the time since the start of the simulation, after the sonic instability reaches saturation, and until the very end, the star supports a strong upper mode with $m$ varying between 12 to 25 at different moments of time (e.g. $m=19$ for $t/2\pi>200$), see Figure \ref{fig:M15.FR.r.a-diag}. In the disk we see a set of global spiral arms, but their number (which evolves in time) does not coincide with the azimuthal wavenumber $m$ of the upper mode present in the star (it is typically lower than $m$); thus, they cannot be the manifestation of a global upper mode in the star-disk system (see \S \ref{sect:vortex} for a discussion of their origin). At later times the arms become less coherent and only one or two reasonably strong arm-like structures are present in the end of the simulation (accompanied by numerous weaker arms).   

Also, after $\sim 275$ orbits the disk exhibits multiple crossing wakes with positive and negative $k_r$, locked inside the resonant cavity near the stellar surface, at $1<r\lesssim 1.17$. Despite the appearance similar to the lower mode in the disk, we interpret these waves to be the high-$\M$ analogues of the low-$m$ modes trapped near the stellar surface that we saw in other runs at lower $M$, see Figs. \ref{fig:M06.HR.r.a} ($m=2$ after 300 orbits for $\M=6$), \ref{fig:M06.HR.r.a} ($m=7$ at 550 orbits for $\M=12$), and \S\ref{sect:low-m_disc}. A notable difference with the lower-$\M$ runs is that for $\M=15$ the azimuthal wavenumber of these wakes is around 20, much higher than at lower $\M$.  
 
A very similar evolutionary sequence is observed in the $\M=13,14$ runs --- persistent dominance of the upper modes inside the star, evolving pattern of the global spirals in the disk, multiple crossing wakes at $r\lesssim 1.2$. The only notable trend that we see is the increase of the azimuthal wavenumber of the observed features with growing $\M$, see \S\ref{sect:dom_modes}. 


\begin{table*}
\begin{tabular}{cccccc}
		{Name} & {$\mathcal{M}$} & {$R_{\rm min}$} & {$N_R$} & {$N_\varphi$} & {Seed type}\\
\hline
\hline
M05.FR.mix.a & 5 & 0.608 & 1024 & 1024 & block-phased-mixed\\
M05.FR.r.a & 5 & 0.608 & 1024 & 1024 & block-random\\
\hline
M06.FR.mix.a & 6 & 0.691 & 1024 & 1024 & block-phased-mixed\\
M06.FR.prime.a & 6 & 0.691 & 1024 & 1024 & prime modes\\
M06.FR.r.a & 6 & 0.691 & 1024 & 1024 & block-random\\
M06.FR.random.a & 6 & 0.691 & 1024 & 1024 & globally random\\
M06.HR.mix.a & 6 & 0.691 & 2048 & 2048 & block-phased-mixed\\
M06.HR.mix.b & 6 & 0.691 & 2048 & 2048 & block-phased-mixed\\
M06.HR.prime.a & 6 & 0.691 & 2048 & 2048 & prime modes\\
M06.HR.r.a & 6 & 0.691 & 2048 & 2048 & block-random\\
M06.HR.r.b & 6 & 0.691 & 2048 & 2048 & block-random\\
M06.HR.r.c & 6 & 0.691 & 2048 & 2048 & block-random\\
\hline
M07.FR.r.a & 7 & 0.752 & 2048 & 2048 & block-random\\
\hline
M08.FR.r.a & 8 & 0.799 & 2048 & 2048 & block-random\\
\hline
M09.LR.mix.a & 9 & 0.834 & 2048 & 2048 & block-phased-mixed\\
M09.LR.mix.b & 9 & 0.834 & 2048 & 2048 & block-phased-mixed\\
M09.LR.prime.a & 9 & 0.834 & 2048 & 2048 & prime modes\\
M09.LR.r.a & 9 & 0.834 & 2048 & 2048 & block-random\\
M09.LR.r.b & 9 & 0.834 & 2048 & 2048 & block-random\\
M09.LR.r.c & 9 & 0.834 & 2048 & 2048 & block-random\\
M09.LR.random.a & 9 & 0.834 & 2048 & 2048 & globally random\\
M09.FR.mix.a & 9 & 0.834 & 4096 & 4096 & block-phased-mixed\\
M09.FR.mix.b & 9 & 0.834 & 4096 & 4096 & block-phased-mixed\\
M09.FR.prime.a & 9 & 0.834 & 4096 & 4096 & prime modes\\
M09.FR.r.a & 9 & 0.834 & 4096 & 4096 & block-random\\
M09.FR.r.b & 9 & 0.834 & 4096 & 4096 & block-random\\
M09.FR.r.c & 9 & 0.834 & 4096 & 4096 & block-random\\
M09.FR.random.a & 9 & 0.834 & 4096 & 4096 & globally random\\
M09.HR.r.a & 9 & 0.834 & 8192 & 8192 & block-random\\
\hline
M10.FR.r.a & 10 & 0.861 & 4096 & 4096 & block-random\\
\hline
M11.FR.r.a & 11 & 0.882 & 4096 & 4096 & block-random\\
\hline
M12.FR.mix.a & 12 & 0.899 & 4096 & 4096 & block-phased-mixed\\
M12.FR.prime.a & 12 & 0.899 & 4096 & 4096 & prime modes\\
M12.FR.r.a & 12 & 0.899 & 4096 & 4096 & block-random\\
M12.FR.r.b & 12 & 0.899 & 4096 & 4096 & block-random\\
M12.FR.random.a & 12 & 0.899 & 4096 & 4096 & globally random\\
\hline
M13.FR.r.a & 13 & 0.913 & 8192 & 8192 & block-random\\
\hline
M14.FR.r.a & 14 & 0.924 & 8192 & 8192 & block-random\\
\hline
M15.FR.r.a & 15 & 0.933 & 8192 & 8192 & block-random\\
\end{tabular}
\caption{See Section~\ref{sect:ic} for descriptions of seed type.}
\label{table:allruns}
\end{table*}

\bsp	
\label{lastpage}
\end{document}